\documentclass[%
reprint,
superscriptaddress,
 amsmath,amssymb,
 aps,
prb,
]{revtex4-2}

\usepackage{graphicx}
\usepackage{dcolumn}
\usepackage{bm}
\usepackage[version=4]{mhchem}
\usepackage{url}
\usepackage{float}
\usepackage[
    colorlinks=true,      
    linkcolor=blue,       
    citecolor=blue,      
    urlcolor=blue         
]{hyperref}
\newcommand{\bk}{\mathbf{k}}
\newcommand{\br}{\mathbf{r}}
\newcommand{\bq}{\mathbf{q}}
\newcommand{\bG}{\mathbf{G}}
\newcommand{\bQ}{\mathbf{Q}}
\newcommand{\bkbar}{\bar{\mathbf{k}}}
\newcommand{\bqbar}{\bar{\mathbf{q}}}

\begin{document}

\title{Artificial moir\'e engineering for an ideal BHZ model}

\author{Wangqian Miao}
\affiliation{Materials Department, University of California, Santa Barbara, California 93106-5050, USA}

\author{Arman Rashidi}
\affiliation{Materials Department, University of California, Santa Barbara, California 93106-5050, USA}

\author{Xi Dai}
\affiliation{Department of Physics, The Hong Kong University of 
Science and Technology, Clear Water Bay, Hong Kong, China}

\date{\today}

\begin{abstract}
We demonstrate that (001) grown \ce{Cd3As2} thin films with a superlattice-patterned gate can potentially realize the moiré Bernevig-Hughes-Zhang (BHZ) model. Our calculations identify the parameterization region necessary to achieve topological flat mini-bands with a $C_{4z}$ symmetric and a $C_{6z}$ symmetric potential. Additionally, we show that a spin-polarized state can serve as the minimal platform for hosting the moir\'e induced quantum anomalous Hall effect, supported by Hartree Fock interaction kernel analysis and self-consistent mean field calculations.
\end{abstract}

\maketitle


\section{Introduction}

Moiré patterns with large periods emerge when layers of van der Waals materials are stacked with a twist or lattice mismatch. These patterns modify energy bands, leading to phenomena such as novel superconductivity \cite{cao2018unconventional,lu2019superconductors,yankowitz2019tuning}, correlated insulators \cite{cao2018correlated,saito2020independent}, and the (fractional) quantum anomalous Hall effect \cite{serlin2020intrinsic, li2021quantum, cai2023signatures, park2023observation, lu2024fractional, xu2023observation, zhang_tbg_qah_2024}. However, inhomogeneities from lattice relaxation and strain are common in these twisted nanodevices, posing significant challenges.

The moiré potential from interlayer coupling is crucial for obtaining flat bands, rather than the twist itself \cite{wan_nearly_2023,gao_untwisting_2023}. A large-period potential that downfolds the original bands and opens gaps at the zone boundary is what matters. These potentials create flat mini-bands, enhancing electron-electron interactions and allowing for engineered band topology.

This introduces the concept of artificial moiré engineering \cite{lian_flat_2020,lu_magic_2024,yang_engineering_2024,ghorashi_topological_2023, wang_dispersion-selective_2024, wang_electrically_2024, anderson_programming_2023, Lu_nc_2023}. By utilizing nanofabrication techniques, we can create designed periodic potentials to modulate the energy bands of two-dimensional electronic systems. Unlike twisted systems, artificial lattices provide flexible control over lattice size, symmetry, and potential strength, offering new degrees of freedom for designing band structures. This approach applies not only to van der Waals multilayers but also to semiconductor thin films grown by molecular beam epitaxy (MBE). One method of creating such an artificial potential is using a twisted hexagonal boron nitride (h-BN) substrate \cite{Kim2023}, where intrinsic charge redistribution generates an external superlattice potential. Additionally, techniques such as dielectric engineering, which involves creating specific patterns during the nanofabrication process, can also be employed for this purpose \cite{rashidi_tuning_2024,sun_signature_2023,wang_dispersion-selective_2024}.

Previous theoretical studies have shown the superlattice potentials can engineer the topology of a single massive Dirac fermion \cite{su_massive_2022,suri_superlattice_2023} while in the case of massless Dirac fermion, a series of satellite Dirac fermions appear \cite{cano_moire_2021, wang_moire_2021}. More recently, people found that the quantum anomalous Hall (QAH) effect is available in Bernal bilayer graphene with a patterned gate \cite{ghorashi_topological_2023, zeng_gate-tunable_2024} when the spin-valley polarized states are favored by the enhanced electron-electron interaction. Beyond these setups based on van der Waals materials, the the effects of superlattice potential on the surface states of the topological insulator (TI) thin films have also been studied in Ref.~\cite{yang_nc_2024}.

In this paper, we primarily discuss the effects of superlattice potentials on a Bernevig-Hughes-Zhang (BHZ) model \cite{bernevig_quantum_2006}, with a special emphasis on (001) grown \ce{Cd3As2} thin films. \ce{Cd3As2}, which belongs to the $D_{4h}$ point group symmetry, was first predicted to be a bulk three-dimensional Dirac semi-metal with strong spin-orbit coupling \cite{wang_cd3as2_prb}. In the thin film region, the topological phase transition from a normal insulator to a two-dimensional topological insulator has been well studied through controlling the thickness in MBE-grown films
\cite{lygo_two-dimensional_2023, guo_zeeman_2023, miao_engineering_2024}, highlighting the effect of quantum confinement. The ability to carry high-density electrons (or holes) and the ease of controlling the band gap make \ce{Cd3As2} thin film an ideal experimental platform to engineer BHZ model, especially compared with other two-dimensional topological insulators \cite{konig_qsh_2007, du_qsh_2015,tang_qsh_wte2_2017}. 

From a theoretical perspective, different from the graphene-based system where valley degree of freedom plays an vital role, the BHZ model itself is simple enough to provide a minimal platform to investigate moir\'e physics. 
The energy bands in the ideal BHZ model are doubly degenerate due to the emergent $\mathcal{PT}$ symmetry, while the spin Chern number or mirror Chern number can still be defined in such cases.  Then it will be intriguing to investigate how the enhanced electron-electron interaction affects the mini-bands that carry non-trivial topology after the superlattice potential is turned on. Specifically, we aim to determine whether the quantum anomalous Hall states can be stabilized as the ground state when only one electron or one hole is doped, driven by the interaction. To address this, we systematically study the behavior of this ideal BHZ model under square and hexagonal superlattice potentials using a two-step strategy. We first present the single-particle topological phase diagram to diagnose the mini-band topology and find the optimal parameter regions where flat mini-bands exist. We then add on the Coulomb interaction and perform mean field calculations. The principle component analysis (PCA) of the Hartree Fock interaction kernel on a two band model shows several modes dominate the spectrum including a spin polarized mode (out-of-plane ferromagnetic state) and two degenerate spin coherent modes (in-plane ferromagnetic state). These magnetic states break time reversal symmetry spontaneously. We further confirm the spin polarized mode can be the ground state with finite Chern number when more adjacent bands are considered.

This paper is organized as follows. In Sec.~\ref{sec:model}, we introduce the model Hamiltonian to describe the (001)-grown \ce{Cd3As2} thin film under a superlattice potential and introduce the mean field technique. In Sec.~\ref{sec:topo_phase}, we present the single-particle topological phase diagram to determine the region where flat topological mini-bands exist. We then study the corresponding correlated phases by analyzing the interaction kernel and performing self-consistent Hartree-Fock calculations, demonstrating that the quantum anomalous Hall phase can be the ground state in such a system in Sec.~\ref{sec:hf_phase}.

\section{Model Hamiltonian and mean field theory \label{sec:model}}

\subsection{BHZ Hamiltonian}

The BHZ Hamiltonian serves as the minimal model to capture the transition from a normal insulator to a two-dimensional topological insulator (TI) \cite{bernevig_quantum_2006, kane_mele_2005}. This Hamiltonian can be decomposed into two spin sectors (Note that, these two sectors can also be labeled with different $\mathcal{M}_z$ eigenvalues), a \(2 \times 2\) sector describing the orbital degree of freedom and its time-reversal conjugate,

\begin{equation}
    H_{\text{BHZ}} = H_0 \oplus \mathcal{T}^{-1} H_0 \mathcal{T} = H_0(\mathbf{k}) \oplus H_0^*(-\mathbf{k}),
\end{equation}
here \(\mathcal{T}\) represents the time reversal operation, and 
\begin{equation}
\begin{aligned}
    H_0 = &m_0 (k_x^2+k_y^2) \sigma_0 + \left(\frac{\delta}{2} + m_1 (k_x^2+k_y^2)\right) \sigma_z  \\
    &+v(k_x \sigma_x - k_y \sigma_y), 
\end{aligned}   
\end{equation}
where \(\delta\), \(m_0\), \(m_1\), and \(v\) are specific material parameters. \(\delta\) characterizes the band gap, and \(\sigma\) represents the Pauli matrices describing the orbital degree of freedom. In the case of (001) grown \ce{Cd3As2} thin film, we take \(m_0=11.5\, \text{eV} \, \text{\AA}^2\), \(m_1=-13.5\, \text{eV} \, \text{\AA}^2\), \(v=0.889\, \text{eV} \, \text{\AA}\) and $\sigma$ represents the degree of freedom of $s$ and $p$ orbitals. These parameters are downfolded from the bulk \(k \cdot p\) model obtained from first-principle calculations \cite{miao_engineering_2024}. The sign of \(\delta\) determines whether the system is in a normal insulator region or a two-dimensional topological insulator region. This phase transition can be controlled by adjusting the thin film thickness in experiments, as inferred from magnetotransport measurements \cite{lygo_two-dimensional_2023, guo_zeeman_2023}. Furthermore, when \(\delta\) and \(m_1\) have different signs, the zeroth Landau Levels (zLLs) become inverted, indicating the system is in the two-dimensional TI phase, see brief discussions in Appendix.~\ref{app:BHZ}. The full BHZ model has full rotational symmetry $C_\infty$ along $z$ direction, inversion symmetry ($\mathcal{P}$), mirror symmetry $\mathcal{M}_z = C_{2z} \mathcal{P}$ and time-reversal symmetry ($\mathcal{T}$) and as a result, the bands are degenerate everywhere. However, the time-reversal symmetry can be manually broken by considering only $H_0$, and the Chern number (mirror Chern number) can be defined for each spin sector ($\mathcal{M}_z$ eigenvalue).

\subsection{Real space moir\'e potential}

\begin{figure}
\includegraphics[width=0.48\textwidth]{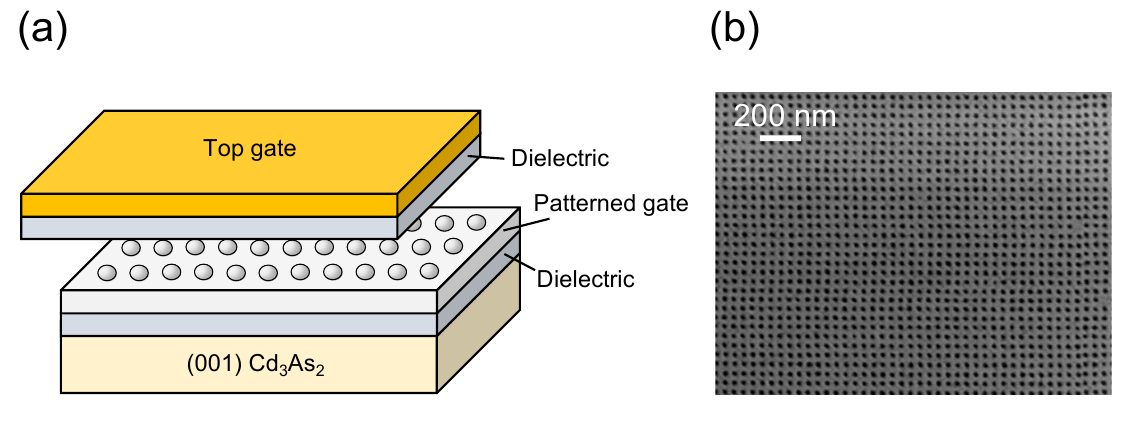}
\caption{\label{fig:exp_setup} A proposed experimental device architecture in which a patterned gate metal creates a moiré potential in the \ce{Cd3As2} thin film. (a) illustration of the double gated device. The patterned gate sets the Fermi level in the film while the top gate controls the potential under the holes, generating an effective moiré superlattice. (b) A scanning electron microscope image of a patterned Ru gate metal realized by electron beam lithography and dry etching. The holes are around 20 nm in diameter while the pitch size is 45 nm.}
\end{figure}

There are several experimental routes for engineering a controllable moiré superlattice potential in an MBE-grown thin film. One method involves patterning the gate metal by etching holes using a dry etch process, followed by a top gate that is separated from the patterned gate by a dielectric, as illustrated in Fig.~\ref{fig:exp_setup}-(a). For instance, ruthenium (Ru) which has an excellent selective dry etching process can be utilized as the patterned gate metal, allowing for the fabrication of small diameter holes with sub-50 nm pitch size \cite{rashidi_tuning_2024}, as demonstrated in the scanning electron microscope image of Fig.~\ref{fig:exp_setup}-(b). In this setup, the patterned gate controls the Fermi level while the top gate modulates the strength of the potential under the holes, effectively forming a moiré superlattice inside the film. We present the general form of a superlattice moiré potential in the real space,
\begin{equation}
  \label{eq:mpot}V(\br) = \sum_{\bG} V(\bG) e^{i \bG \cdot \br}  
\end{equation}
where $\bG$ is the moir\'e reciprocal lattice and obviously this potential obeys the translational symmetry of the moir\'e period. This potential is a scalar potential that does not couple with the spin and orbital degrees of freedom. This is a good approximation in the case of (001) grown high-quality \ce{Cd3As2} thin films where surface states do not exist. However, in the case of Bernal bilayer graphene \cite{zeng_gate-tunable_2024} and the \ce{Bi2Se3} thin film family \cite{cano_moire_2021,yang_nc_2024}, the moiré potential exhibits layer dependence, making this scalar potential an oversimplification of the real scenario. In this paper, we mainly focus on the $C_{6z}$ symmetric and $C_{4z}$ symmetric potentials with one shell of $\bG$ vectors. For the $C_{6z}$ case, only  $\bG_1 = \frac{4\pi}{\sqrt{3}L} \left(\frac{1}{2}, -\frac{\sqrt{3}}{2}\right)$, $\bG_2 = \frac{4\pi}{\sqrt{3}L} \left(\frac{1}{2}, \frac{\sqrt{3}}{2}\right)$, $\bG_3 = -\bG_1 - \bG_2$, $-\bG_1$, $-\bG_2$ and $-\bG_3$ are preserved in the summation in Eq.~\ref{eq:mpot}. For the $C_{4z}$ case, only $\bG_1 = \frac{2\pi}{L} (1, 0)$, $\bG_2 = \frac{2\pi}{L} (0, 1)$, $-\bG_1$ and $-\bG_2$ are preserved. The length of the real space potential $L$ is set at 50 nm in this paper, aligning with recent experiments \cite{rashidi_tuning_2024}. Now, the full Hamiltonian considering the effect of superlattice potential should be,
\begin{equation}
 H = H_{\text{BHZ}} (\bk\rightarrow -i \partial_\br) + V(\br),  
\end{equation}
This can be solved under a series of plane waves and we set the cutoff of the plane waves to be $5|\bG|$ in our calculation.

\subsection{Hartree Fock mean field theory}
In the moir\'e flat band systems, the bandwidth of the mini-bands is comparable with the interaction strength and this suggests the effect from Coulomb interaction cannot be ignored. 
Here, we present the standard band projected Hartree Fock mean field theory \cite{nick_prx_2020, zhang_hf_2020, zhang_citbg_prl_2022} used to study the Coulomb interactions in the moir\'e systems and introduce the concept of the Hartree Fock interaction kernel.

The interacting part of the Hamiltonian can be written in the plane wave basis,
\begin{equation}
    H_{\text{int}} = \frac{1}{2 N_\bk} \sum_{\bk,\bk',\bq}\sum_{\mu, \mu'} v_{\bq}c_{\bk+\bq,\mu}^\dagger c_{\bk'-\bq,\mu'}^\dagger c_{\bk',\mu'} c_{\bk,\mu},
\end{equation}
where $\mu=(\tau,s)$ is the joint index for orbital and spin. Moreover, we adopt the gate screened interaction,
$v_{\bq} = \frac{e^2}{2 \epsilon_0\epsilon \Omega_s} \frac{\tanh (d_s q)}{q}$
where $q=|\bq|$, $\Omega_s$ is the size of the moir\'e unit cell and $\epsilon_0$ is the permittivity of vacuum, $\epsilon$ is the relative permittivity to control the strength of the interaction and the gate distance $d_s = 10$ nm is fixed in our later calculations. 

To prevent from the expensive cost of performing self-consistent calculations in the original plane wave basis, we project the Hamiltonian into the active bands subspace (denoted by indices $mnm'n'$), the interacting part can be written as,
\begin{widetext}
\begin{equation}
    \begin{aligned}
    H_{\text{int}} =\frac{1}{2N_{\bkbar}} \sum_{\bkbar \bkbar' \bqbar} \sum_{\bQ} \sum_{mnm'n'}v(\bqbar+\bQ)& \lambda_{mn}(\bkbar,\bqbar,\bQ) \lambda_{m'n'}(\bkbar',-\bqbar,-\bQ) 
    d_m^\dagger(\bkbar+\bqbar)d_{m'}^\dagger(\bkbar'-\bqbar)d_{n'}(\bkbar') d_n(\bkbar).
    \end{aligned}
\end{equation}
\end{widetext}
where we denote $\bG=(\bG,\mu)$, $\bkbar,\bqbar$ as the momentum in the mini-Brillouin zone (mB.Z.) and the form factor $\lambda$ represents for,
\begin{equation}
    \lambda_{mn}(\bkbar,\bar{\bq},\bQ) = \sum_\bG D_{\bG+\bQ,m}^*(\bkbar+\bqbar)D_{\bG,n}(\bkbar),
\end{equation}
$D_{\bG,n}$ is the eigen wavefunction for the $n$-th band. Under Hartree Fock mean field approximation, the prefactors before the four-fermion operators can be divided into two parts,
\begin{equation}
    \begin{aligned}
    H_{mnm'n'} (\bkbar, \bkbar') =& \sum_{\bQ} v(\bQ) \lambda_{mn}(\bkbar,\mathbf{0},\bQ) \lambda_{m'n'}(\bkbar',\mathbf{0},-\bQ)\\
    F_{m'n m n'}(\bkbar, \bkbar') =& \sum_{\bQ} v(\bkbar'-\bkbar+\bQ) 
        \lambda_{mn}(\bkbar, \bkbar'-\bkbar,\bQ) \\
        &\lambda_{m'n'}(\bkbar, \bkbar-\bkbar',-\bQ).     
    \end{aligned}
\end{equation}
The first one corresponds to the Hartree term and the second one is the Fock term. The $m,m'$ subscript in the Fock term is exchanged due to the normal ordering. The total interacting tensor now can be defined as,
\begin{equation}
    V_{mn m'n'}(\bkbar,\bkbar') = H_{mnm'n'}(\bkbar,
    \bkbar')-F_{m'nmn'}(\bkbar,\bkbar').
\end{equation}
The Hartree Fock correction for the original Hamiltonian is a two-fermion term,
\begin{equation}
    \label{eq:hmf}
    H_{\mathrm{hf}} = \frac{1}{N_{\bkbar}} \sum_{mn\bkbar}\sum_{m'n'\bkbar'} V_{mnm'n'}(\bkbar,\bkbar') \rho_{m'n'}(\bkbar') d_{m}^\dagger(\bkbar) d_{n}(\bkbar).
\end{equation}
where the density matrix defined as following is the order parameter,
\begin{equation}
    \rho_{m'n'}(\bkbar) = \left< d_{m'}^\dagger(\bkbar') d_{n'}(\bkbar')\right>.
\end{equation}
The mean filed Hamiltonian can be solved self consistently until the order parameters are converged. The criterion for convergence of the self-consistent calculations is that the difference between the updated and previous density matrices is less than $10^{-9}$ ($\sum_{\mathbf{k}} ||\rho_{i+1}(\mathbf{k}) - \rho_{i}(\mathbf{k})|| < 10^{-9}$). The energy functional of the system is written as,
\begin{equation}
\begin{aligned}
    \label{eq:en_func}
    E[\rho] =& \sum_{\bkbar mn} h^0_{mn}(\bkbar)\rho_{mn}(\bkbar) + 
    \frac{1}{2N_{\bkbar}} \sum_{mn\bkbar}\sum_{m'n'\bkbar'} \\ 
    & \rho_{mn}(\bkbar)V_{mnm'n'}(\bkbar,\bkbar') \rho_{m'n'}(\bkbar') .
\end{aligned}
\end{equation}
The second part of Eq.~\ref{eq:en_func} denoted as $E_\text{int}[\rho]$ is the interaction part of the energy functional and the prefactor 1/2 is used to deduct the condensation energy. After expanding the density matrix using the generalized Pauli matrices $\Gamma_{mn}^\gamma$. We can define the Hartree Fock kernel,
\begin{equation}
    K_{\gamma \bk, \gamma' \bk'} = 
    \frac{1}{2N_{\bkbar}} 
    \sum_{mnm'n'} \Gamma_{mn}^\gamma V_{mnm'n'}(\bkbar,\bkbar') \Gamma_{m'n'}^{\gamma'}.
\end{equation}
which is real and symmetric. Then, it is natural to perform spectrum analysis and find the corresponding leading modes,
\begin{equation}
\begin{aligned}
        \label{eq:kernel_decomp}
        K_{\gamma \bk, \gamma' \bk'} &= \sum_\ell K_\ell C_{\gamma \bk, \ell}  C_{\ell, \gamma' \bk'} \\
        &\approx \sum_{i} K_i C_{\gamma \bk, i}  C_{i, \gamma' \bk'}
\end{aligned}
\end{equation}
We denote \( C_{\gamma \bk, \ell} \) as the \(\ell\)-th eigenvector of \( K_{\gamma \bk, \gamma' \bk'} \), with \( K_\ell \) being the corresponding eigenvalue. From now on, we refer to \( C_{\gamma \bk, \ell} \) as the \(\ell\)-th eigenmode (or eigenchannel) of the Hartree-Fock kernel. Typically, several modes serve as the principal components of this kernel, suggesting that we only need to consider these modes rather than preserving all modes, as in the original band-projected Hartree-Fock procedure. In the second step of Eq.~\ref{eq:kernel_decomp}, \( i \) represents the index for the principal components, and the ratio between \(\sum_i K_i \) and  \( \sum_\ell K_\ell \) can be the criterion to  determine whether this principal component analysis (PCA) is effective in a specific case.

\begin{figure}
\includegraphics[width=0.5\textwidth]{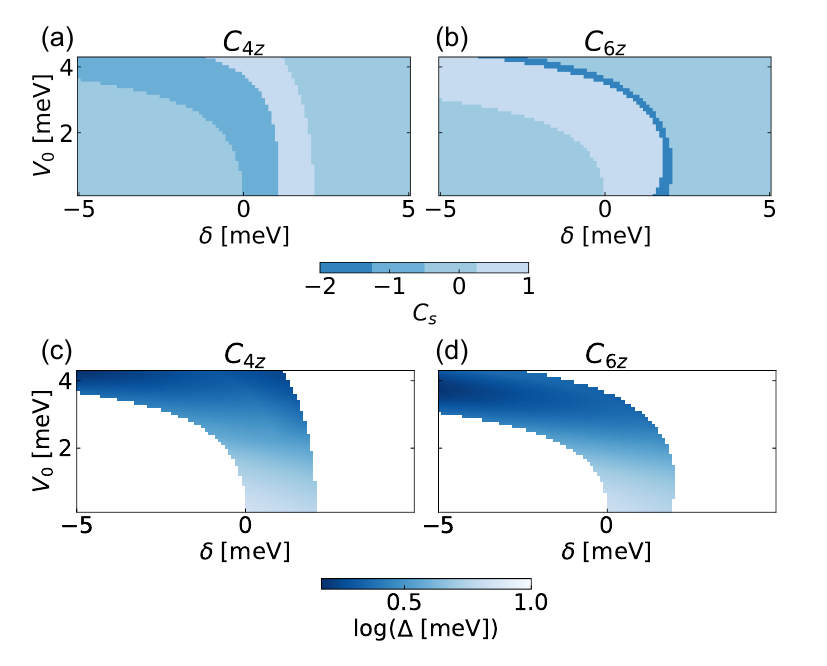}
\caption{\label{fig:topo_phase} Topological phase diagram of the moir\'e BHZ model. (a) The calculated spin Chern number ($C_s$) of the first valence bands when a $C_{4z}$ symmetric potential is applied, $\delta$ characterize the original band gap and $V_0$ is the strength of moir\'e potential. (b) Same as (a) but with a $C_{6z}$ symmetric moir\'e potential. (c) The bandwidth $\Delta$ of the first valence bands in the topological region when a $C_{4z}$ symmetic potential is applied. (d) Same as (c) but with a $C_{6z}$ symmetric moir\'e potential.
}
\end{figure}

\section{Single particle topological phase diagram \label{sec:topo_phase}}

The non-trivial topology of the moiré BHZ model, previously discussed in a recent work using the method of topological quantum chemistry \cite{yang_engineering_2024}, demonstrates the presence of non-trivial mini-band topology. In this section, we focus on real material parameters and investigate whether topological flat bands are available in the system of \ce{Cd3As2}, which exhibits strong particle-hole asymmetry not considered in Ref.~\cite{trithep_2024}.

\begin{figure}
\includegraphics[width=0.48\textwidth]{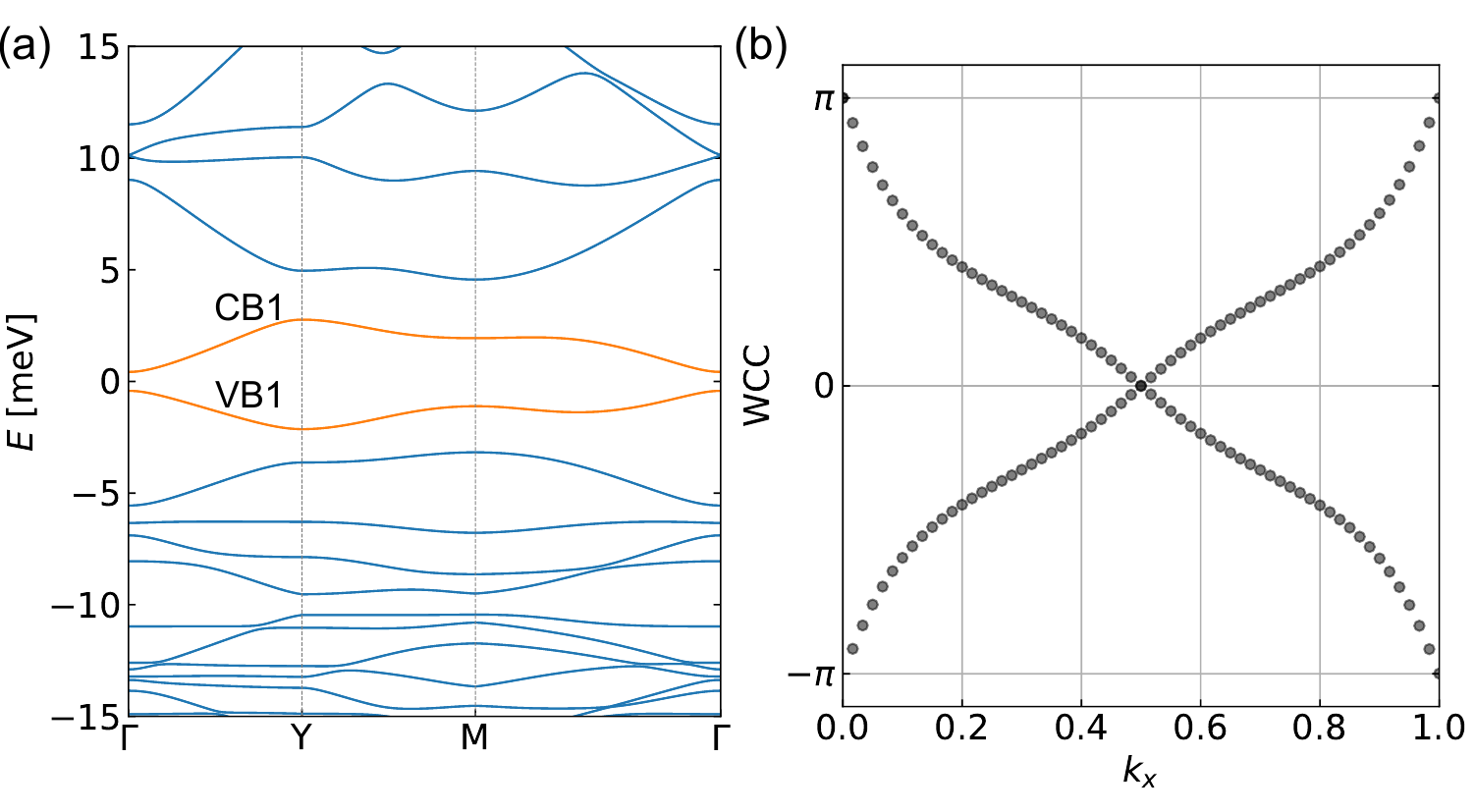}
\caption{\label{fig:flat_band_c4} (a) Single particle band structures for the moir\'e BHZ model when a $C_{4z}$ symmetric potential is applied ($\delta=-5$ meV, $V_0$ = 4 meV). The first valence bands and the first conduction bands are labeled and are relatively flat. (b) Wannier charge center (WCC) flow of the first valence bands (VB1). The non-trivial winding feature 
indicates the band topology which is consistent with the spin Chern number calculation.}
\end{figure}

Our calculations primarily examine the first valence bands of the system when a weak moiré potential is applied (meaning the strength of the potential $V_0$ is comparable to the magnitude of the original band gap $\delta$). In Fig.~\ref{fig:topo_phase}, we present the calculated spin Chern number $C_s$ under a $C_{4z}/C_{6z}$ symmetric potential. A large area in the phase diagram is topologically non-trivial, carrying $C_s = \pm1, 2$. We also examine the bandwidth of the mini-bands and find that when the strength of the potential $V_0$ is approximately equal to the magnitude of the original gap $\delta$ in the normal region, there is an optimal chance to create topological flat bands. This is confirmed in our band structure calculations. The typical single-particle energy spectrum for a moiré BHZ model is shown in Fig.~\ref{fig:flat_band_c4}. Here, we consider a specific case with a $C_{4z}$ symmetric potential. The band gap is set to be $\delta= -5$ meV and the moir\'e potential strength is $V_0 = 4$ meV. The non-trivial winding of the Wannier charge center (WCC) of the first valence band is consistent with previous spin Chern number calculations, confirming the topological nature. This valence band has a bandwidth around 2 meV while the electron-electron interaction can also be estimated through $e^2/4\pi\epsilon \epsilon_0 L \approx 3$ meV when $\epsilon=10$, $L=50$ nm. Then the effect from interaction cannot be ignored.

\section{Correlated phases and mean field phase diagram \label{sec:hf_phase}}

\begin{figure}
\includegraphics[width=0.48\textwidth]{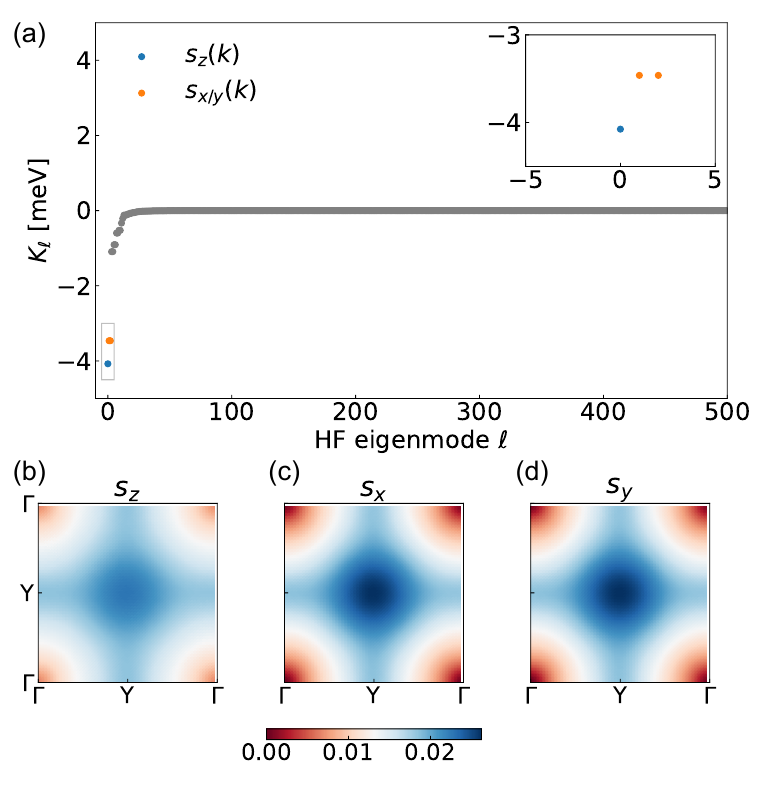}
\caption{\label{fig:hf-mode} (a) Eigenvalues of different modes of the Hartree Fock interaction kernel. Three discrete modes are separated from others, correspond to one spin polarized mode ($s_z (\bk)$, labeled in blue) and two spin coherent modes ($s_{x/y}(\bk)$, labeled in orange). The single particle Hamiltonian is solved on a $40\times 40$ $k$-mesh when a $C_{4z}$ symmetric potential is applied ($\delta=-5$ meV, $V_0$ = 4 meV) (b-c) The $k$ space distribution of the Hartree Fock eigenmodes for the spin-polarized mode and two degenerate spin-coherent modes.}
\end{figure}

We begin by discussing the potential order parameters stabilized by Coulomb interactions, analyzed through the eigenmodes of the Hartree-Fock interaction kernel. Our discussion is confined to the flat band subspace via band projection, specifically the degenerate first valence band subspace. Here, we emphasize that this two-band system gives a minimal example to show the interaction-induced topologically non-trivial phases. In the strong coupling limit of the flat band systems, we can directly inferring the properties of the system by diagnosing the interaction kernel because Coulomb interactions dominate and the kinetic part can be ignored \cite{ashvin_matbg_prl_2019}. The interaction kernel analysis provides us the potential orders which can spontaneously form to lower down the total energy in the system. Taking the $C_{4z}$ symmetric case stated in Sec.~\ref{sec:topo_phase} as an example, the eigenvalues of different modes and their $k$ space distributions are shown in Fig.~\ref{fig:hf-mode}. Three discrete modes exhibit lower energy and are well separated from other continuous ones. The first mode, highlighted by blue dot, corresponds to the spin-polarized (SP) mode (out-of-plane ferromagnetic state) $s_{z}(\bk)$ , while the two degenerate modes, labeled in orange, correspond to the spin-coherent modes (in-plane ferromagnetic state) $s_{x/y}(\bk)$. This indicates that Coulomb interactions favor these magnetic orders which spontaneously break time reversal symmetry. Notably, the SP mode can also induce a quantum anomalous Hall effect.
The same analysis can be done for the $C_{6z}$ symmetric case, which is shown in Appendix.~\ref{app:c6}.

\begin{figure}
\includegraphics[width=0.48\textwidth]{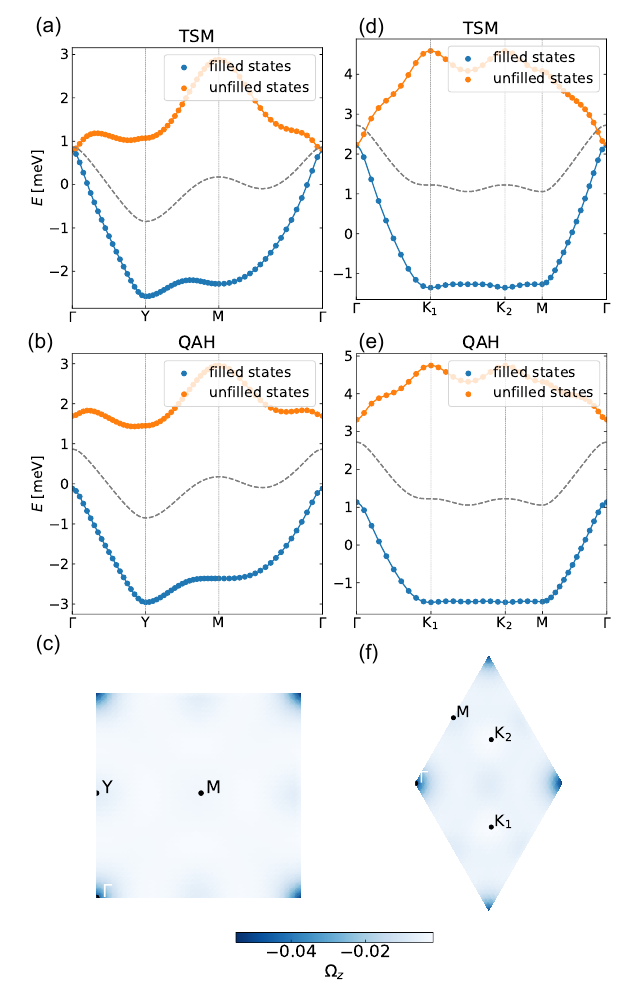}
\caption{\label{fig:2band_hf} Two band Hartree Fock calculations when a $C_{4z}$ symmetric potential ($\delta=-5$ meV, $V_0$ = 4 meV) is applied and a $C_{6z}$ symmetric potential ($\delta=-5$ meV, $V_0$ = 3.5 meV) is applied. The self-consistent Hartree Fock calculations are performed on a 40 $\times$ 40 $k$-mesh for the $C_{4z}$ symmetric case and on a 30 $\times$ 30 $k$-mesh for the $C_{6z}$ symmetric case. (a), (d) Hartre -Fock band structures for the spin-coherent state, exhibiting a topological semimetal (TSM) phase. The original bands are labeled in grey dashed line. (b), (e) Hartree Fock band structures for the spin-polarized state, indicative of a quantum anomalous Hall (QAH) phase. (c), (f) Berry curvature distribution in the moiré Brillouin zone (mBZ), with labeled high-symmetry points. }
\end{figure}

\begin{figure*}
\includegraphics[width=0.88\textwidth]{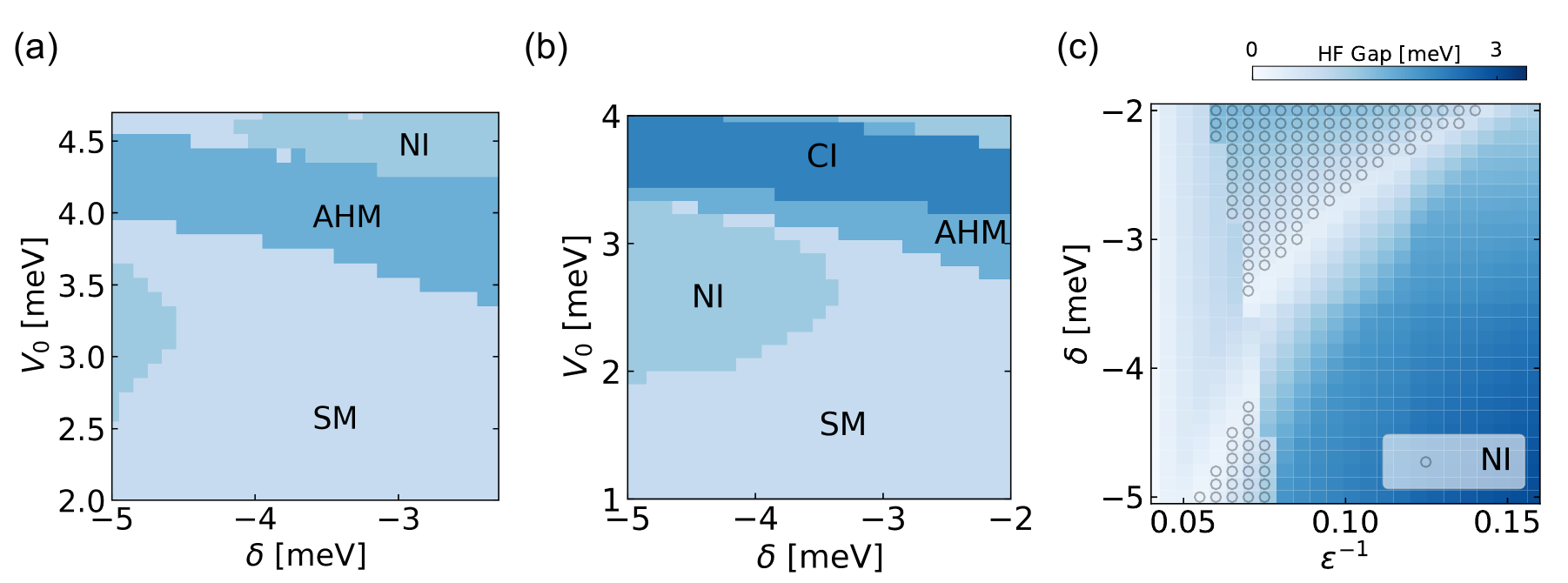}
\caption{\label{fig:hf_phase} Mean field phase diagram for the \ce{Cd3As2} thin films under superlattice potential (a) The original band gap $\delta$ and moir\'e potential strength are set as parameters and the relative permittivity $\epsilon$ is set as 25 when a $C_{4z}$ symmetric potential applied. The Chern insulator (CI) phase can not be stabilized but the anomalous Hall metal phase (AHM) appears. SM denotes the semi-metal phase and NI denotes the normal insulator phase. (b) same as (a) but with a $C_{6z}$ symmetric potential. The Chern insulator phase can be stabilized in this case.  (c) We investigate the effect of interaction strength in the optimal region of panel (b), by setting $V_0=3.5$ meV. The original band gap $\delta$ and the relative permittivity $\epsilon$ are set as tuning parameters. The increasing of interacting strength will result in band mixing and topological phase transition happens. }
\end{figure*}

We then perform self-consistent Hartree-Fock calculations using a projected two-band model under $C_{4z}$ or $C_{6z}$ symmetric potentials, starting with an initial density matrix directly from the eigenmode of the interaction kernel. The corresponding $k$ space distribution is shown in the second row of Fig.~\ref{fig:hf-mode}. The chemical potential was adjusted to fill one hole per moiré unit cell. The results converged into two distinct phases, as shown in Fig.~\ref{fig:2band_hf}. The first row depicts a topological semimetal phase that breaks time-reversal while preserving $C_{2z}\mathcal{T}$ symmetry. This phase exhibits spin coherence and corresponds to the order parameter $s_{x/y}(\bk)$. Notably, the node at the $\Gamma$ point remains ungapped by Coulomb interactions, similar to features recently discussed in magic angle twisted bilayer graphene (MATBG) when considering electron-optical phonon coupling \cite{shang_nematic_tsm_2021, shi_epc_tbg_2024, wang_epc_tbg_2024}. This feature is consistent with the eigenmode analysis which shows $s_{x/y} (\Gamma) = 0$ in Fig.~\ref{fig:hf-mode}-(c), (d). The second row presents an insulating phase that spontaneously breaks time-reversal and $C_{2z}\mathcal{T}$ symmetries. This phase displays a spin-polarized feature, confirmed by examining the wavefunction components, and corresponds to the order parameter $s_z(\bk)$. Note that, $s_z(\Gamma)$ is not equal to zero in the eigenmode analysis which explain the gap opened at $\Gamma$ point. Interestingly, the bands carry non-trivial Chern numbers of $\pm 1$, indicating a quantum anomalous Hall insulator phase, confirmed by integrating the Berry curvature, as shown in the third row of Fig. \ref{fig:2band_hf}. Both interacting phases benefit energetically from spontaneous symmetry breaking upon hole doping. Furthermore, the valence bands may invert with adjacent bands and change the topological property when the band gap is small compared with the bandwidth and interaction strength. Determining the true ground state requires extending the mean-field calculations to include more bands.

We now conduct a more rigorous multi-band self-consistent Hartree Fock calculation to verify our previous analysis. The full picture of the Hartree-Fock mean-field phase diagram is shown in Fig.~\ref{fig:hf_phase}. This diagram is calculated at zero temperature on a $10 \times 10 $ $k$-mesh for the $C_{4z}$ case and a $15 \times 15$ $k$-mesh for the $C_{6z}$ case by projecting the full Hamiltonian onto six bands (four valence bands and two conduction bands) to fully account for band mixing effects. The chemical potential is set to one hole per moir\'e unit cell. We first fix $\epsilon = 25$, indicating that the interaction is at an intermediate level. The phase diagram is shown in panel (a) and (b) of Fig.~\ref{fig:hf_phase}. We found that the Chern insulator phase ($|\mathcal{C}|=1$) with order parameters $\sigma_0 s_z, \sigma_z s_z$ can be stabilized in the $C_{6z}$ case by tuning the band gap $\delta$ and the moiré potential strength $V_0$ (The optimal region is $V_0 \approx$ 4 meV and $-5<\delta<-2$ meV). Furthermore, the anomalous Hall metal (AHM) phase can also appear, indicating no global band gap in the mB.Z. while the bands carry non-trivial Berry curvature. The typical spin-resolved Hartree Fock band structures in the mean field phase diagram are shown in Appendix.~\ref{app:hf_band}. We study how interaction strength affects the Chern insulator phase in panel (c) when $C_{6z}$ symmetric potential is applied. Here, the moiré potential strength is fixed at 3.5 meV, and the relative permittivity $\epsilon$ is used as a tuning parameter. The normal insulator (NI) phase is labeled with white circles, while other areas in the phase diagram are in the CI phase. Increasing interaction strength initially causes band mixing, pushing the Chern insulator phase into a normal insulator phase. However, with further increase in interaction strength ($\epsilon \approx$ 10), the CI phase can be reentrant, and the HF band gap can be enhanced to around 3 meV.

\section{Conclusion}

In summary, we utilize (001) \ce{Cd3As2} thin films with patterned gates to realize the moiré BHZ model. Unlike twisted bilayer materials, where the moiré potential arises from interlayer interactions and mechanical modulations like lattice relaxation and strain, our material remains uniform and clean under a static real-space electrical potential. By carefully designing the superlattice potential and tuning the system's band gap, we achieve topological flat minibands and confirm that the quantum anomalous Hall phase can be the ground state through hole doping in this region. Looking ahead, it would be intriguing to explore whether such long-period modulation can spontaneously occur without patterned gates when the system is slightly doped, potentially forming a Wigner crystal or anomalous Hall crystal \cite{sheng_qah_prl_2024,zeng_topo_ahc_2024,dong_ahc_2023, dong_ahc_senthil_2024, tan_parent_bc_2024}. It will also be interesting to examine how the density wave orders \cite{kwan_prb_2021,jihang_dw_2024} in the two-dimensional electron gas compete with these ferromagnetic orders, which we leave for future research. The ability to control and manipulate these topological states in artificial moiré lattices opens up new possibilities beyond twisting.

\begin{acknowledgments}
We acknowledge Prof. Susanne Stemmer for telling us recent progress on the superlattice gating on \ce{Cd3As2} thin films. W. Miao acknowledges fruitful discussions with H. Shi, T.Y. Qiao, B.H. Guo, B.L. Liu and T. Devakul. X. Dai is supported by a fellowship award from the Research Grants Council of the Hong Kong Special Administrative Region, China (Project No. C7037-22GF). The authors thank \texttt{GNU Parallel} \cite{gnu_parallel} used in the Hartree Fock code.
\end{acknowledgments}

\appendix

\section{Details of the BHZ model \label{app:BHZ}}

\begin{figure}
\includegraphics[width=0.48\textwidth]{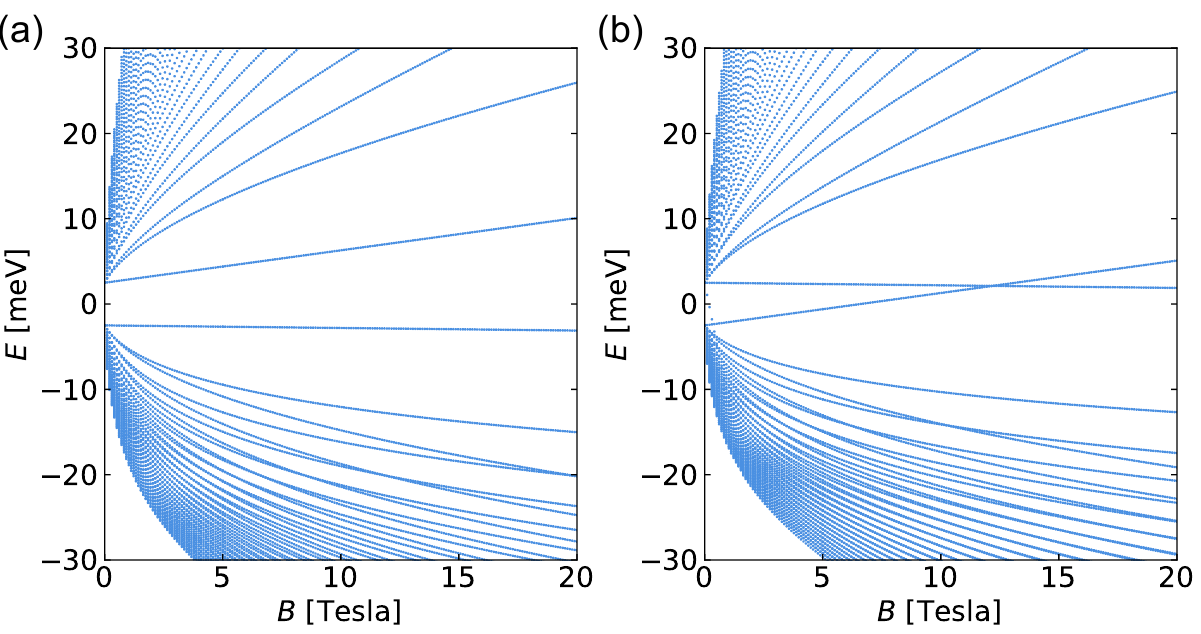}
\caption{\label{fig:bhz_LL} The Landau level diagram calculated using the BHZ model proposed for \ce{Cd3As2}, using parameters defined in Sec.~\ref{sec:model} (a) $\delta=-5$ meV, normal region, the zLLs are not inverted. (b) $\delta=5$ meV, 2D TI region, the zLLs are inverted. This corresponds to the band inversion between $s$ and $p$ band happened at $\Gamma$ point.} 
\end{figure}

The band gap and the corresponding topological feature ($Z_2$ index) can be inferred from the Landau fan diagram in magnetotransport measurements for \ce{Cd3As2} thin film. We calculated the Laudau level diagram using the Hamiltonian defined in Sec.~\ref{sec:model} in the main text. When the sign of band gap is negative $\delta<0$, the system is in the normal region, the zeroth Landau levels (zLLs) are not inverted. When the sign of the band gap is positive $\delta>0$, the system is in the two-dimensional topological insulator (2D TI) phase, e.g., the zLLs are inverted, as shown in Fig.~\ref{fig:bhz_LL}. The calculated result is consistent with recent experiment \cite{lygo_two-dimensional_2023}.

\section{Hartree Fock eigenmodes of the $C_{6z}$ symmetic case \label{app:c6}}

\begin{figure}
\includegraphics[width=0.48\textwidth]{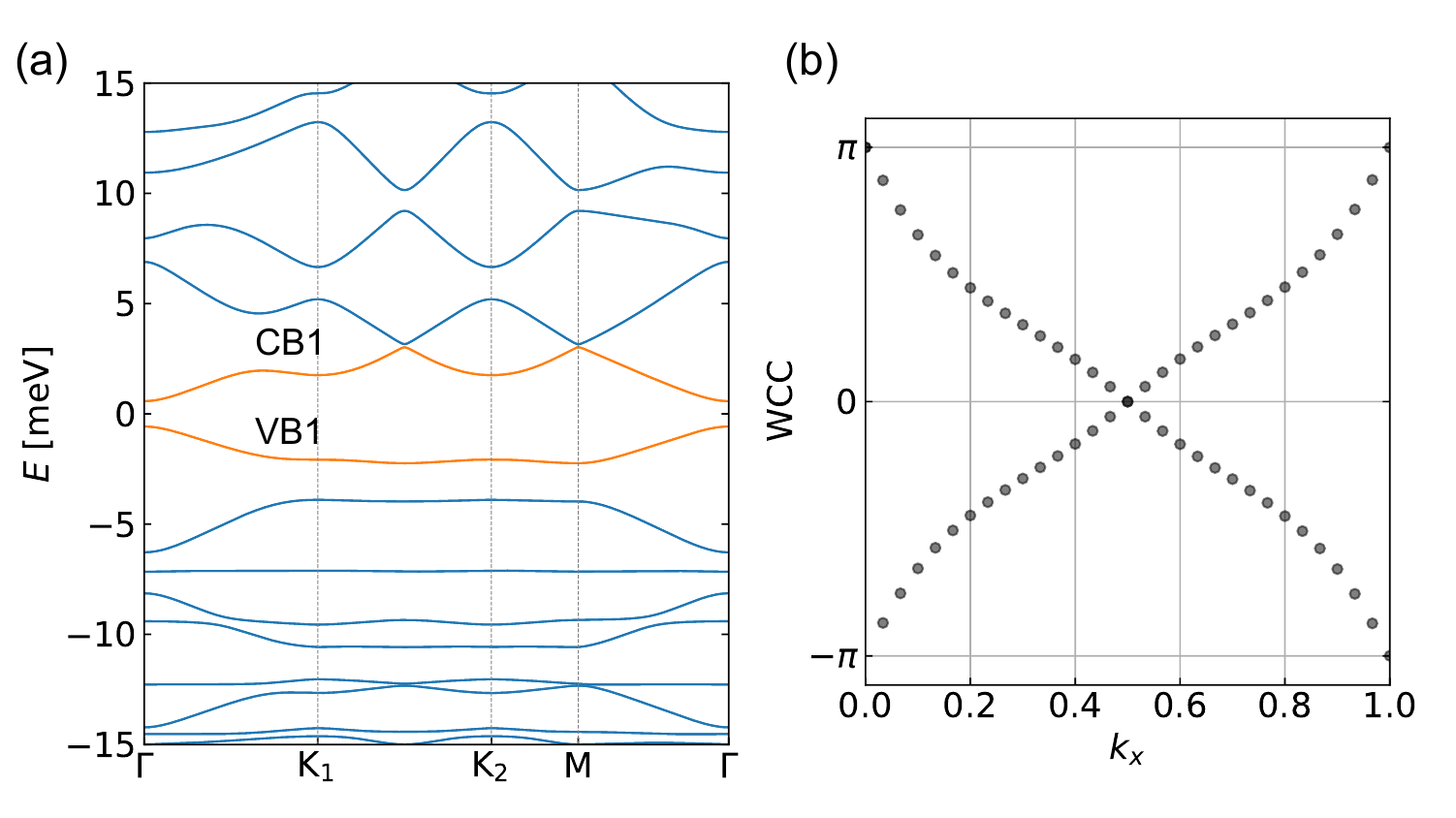}
\caption{\label{fig:flat_band_c6} (a) Single particle band structures for the moir\'e BHZ model when a $C_{6z}$ symmetric potential is applied ($\delta=-5$ meV, $V_0$ = 3.5 meV). The first valence bands and the first conduction bands are labeled and are relatively flat. (b) Wannier charge center (WCC) flow of the first valence bands. The non-trivial winding feature 
indicates the band topology which is consistent with the spin Chern number calculation.}
\end{figure}

\begin{figure} 
\includegraphics[width=0.48\textwidth]{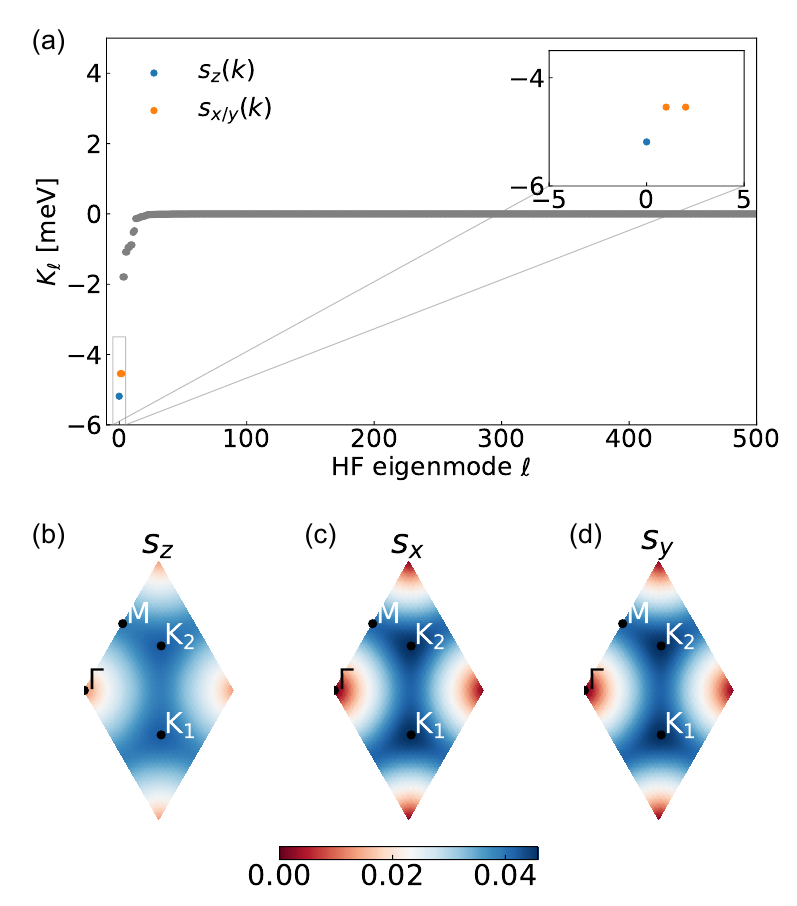}
\caption{\label{fig:hf-mode-c6} (a) Eigenvalues of different modes of the Hartree Fock interaction kernel. Three discrete modes are separated from others, correspond to one spin polarized mode ($s_z{\bk}$, labeled in blue) and two spin coherent modes ($s_{x/y}(\bk)$, labeled in orange). The single particle Hamiltonian is solved on a $30\times 30$ $k$-mesh when a $C_{6z}$ symmetric potential is applied ($\delta=-5$ meV, $V_0$ = 3.5 meV) (b-c) The $k$ space distribution of the Hartree Fock eigenmodes for the spin-polarized mode and two degenerate spin-coherent modes.}
\end{figure}

We show the typical single particle spectrum in Fig.~\ref{fig:flat_band_c6} when a $C_{6z}$ symmetric potential is applied and the corresponding Hartree fock eigenmodes in Fig.~\ref{fig:hf-mode-c6}. The flat band feature and the discrete eigenmodes share similarities as the case of $C_{4z}$ symmetric case discussed in Sec.~\ref{sec:hf_phase}.

\section{Spin resolved Hartree Fock band structures \label{app:hf_band}}

\begin{figure}[!h]
\includegraphics[width=0.48\textwidth]{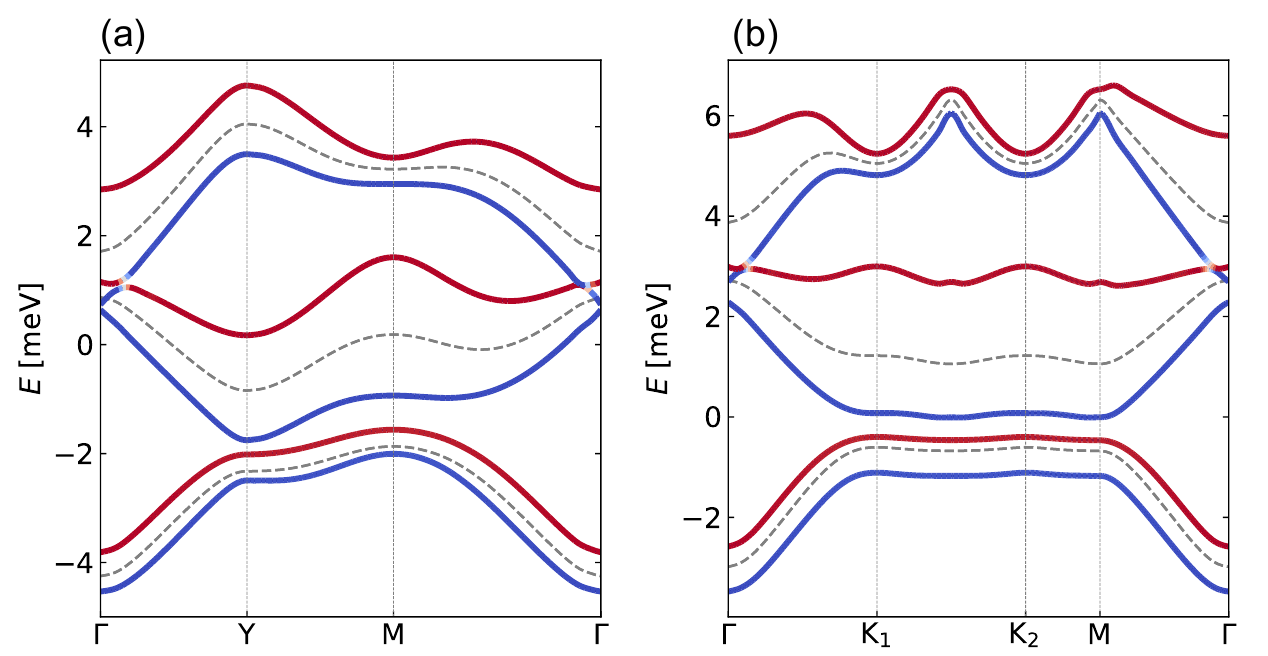}
\caption{\label{fig:hf-6band} (a) Six-band Hartree Fock band structures performed on a $20 \times 20$ $k$-mesh with a $C_{4z}$ symmetric potential. ($\delta=-5 \,\text{meV}, V_0 = 4 \,\text{meV}, \epsilon = 25$) This corresponds to a anomalous Hall metal phase (AHM). (b) Six-band Hartree Fock band structures performed on a $18 \times 18$ $k$-mesh with a $C_{6z}$ symmetric potential ($\delta=-5 \,\text{meV}, V_0 = 3.5 \,\text{meV}, \epsilon = 25$). This corresponds to a quantum anomalous Hall phase (QAH).}
\end{figure}

In this appendix, we present the spin resolved Hartree Fock band structures in Fig.~\ref{fig:hf-6band} by projecting the Hamiltonian onto six bands, including four valence bands and two conduction bands. Panel (a) depicts the anomalous Hall metal phase (AHM) and panel (b) depicts the quantum anomalous Hall (QAH) insulator phase. The blue color labels the `spin up' states while the red color labels the `spin down' states. Both occupied states show the spin-polarized features.

\newpage
\bibliography{apssamp}

\begin{thebibliography}{56}%
\makeatletter
\providecommand \@ifxundefined [1]{%
 \@ifx{#1\undefined}
}%
\providecommand \@ifnum [1]{%
 \ifnum #1\expandafter \@firstoftwo
 \else \expandafter \@secondoftwo
 \fi
}%
\providecommand \@ifx [1]{%
 \ifx #1\expandafter \@firstoftwo
 \else \expandafter \@secondoftwo
 \fi
}%
\providecommand \natexlab [1]{#1}%
\providecommand \enquote  [1]{``#1''}%
\providecommand \bibnamefont  [1]{#1}%
\providecommand \bibfnamefont [1]{#1}%
\providecommand \citenamefont [1]{#1}%
\providecommand \href@noop [0]{\@secondoftwo}%
\providecommand \href [0]{\begingroup \@sanitize@url \@href}%
\providecommand \@href[1]{\@@startlink{#1}\@@href}%
\providecommand \@@href[1]{\endgroup#1\@@endlink}%
\providecommand \@sanitize@url [0]{\catcode `\\12\catcode `\$12\catcode `\&12\catcode `\#12\catcode `\^12\catcode `\_12\catcode `\%12\relax}%
\providecommand \@@startlink[1]{}%
\providecommand \@@endlink[0]{}%
\providecommand \url  [0]{\begingroup\@sanitize@url \@url }%
\providecommand \@url [1]{\endgroup\@href {#1}{\urlprefix }}%
\providecommand \urlprefix  [0]{URL }%
\providecommand \Eprint [0]{\href }%
\providecommand \doibase [0]{https://doi.org/}%
\providecommand \selectlanguage [0]{\@gobble}%
\providecommand \bibinfo  [0]{\@secondoftwo}%
\providecommand \bibfield  [0]{\@secondoftwo}%
\providecommand \translation [1]{[#1]}%
\providecommand \BibitemOpen [0]{}%
\providecommand \bibitemStop [0]{}%
\providecommand \bibitemNoStop [0]{.\EOS\space}%
\providecommand \EOS [0]{\spacefactor3000\relax}%
\providecommand \BibitemShut  [1]{\csname bibitem#1\endcsname}%
\let\auto@bib@innerbib\@empty
\bibitem [{\citenamefont {Cao}\ \emph {et~al.}(2018{\natexlab{a}})\citenamefont {Cao}, \citenamefont {Fatemi}, \citenamefont {Fang}, \citenamefont {Watanabe}, \citenamefont {Taniguchi}, \citenamefont {Kaxiras},\ and\ \citenamefont {Jarillo-Herrero}}]{cao2018unconventional}%
  \BibitemOpen
  \bibfield  {author} {\bibinfo {author} {\bibfnamefont {Y.}~\bibnamefont {Cao}}, \bibinfo {author} {\bibfnamefont {V.}~\bibnamefont {Fatemi}}, \bibinfo {author} {\bibfnamefont {S.}~\bibnamefont {Fang}}, \bibinfo {author} {\bibfnamefont {K.}~\bibnamefont {Watanabe}}, \bibinfo {author} {\bibfnamefont {T.}~\bibnamefont {Taniguchi}}, \bibinfo {author} {\bibfnamefont {E.}~\bibnamefont {Kaxiras}},\ and\ \bibinfo {author} {\bibfnamefont {P.}~\bibnamefont {Jarillo-Herrero}},\ }\bibfield  {title} {\bibinfo {title} {Unconventional superconductivity in magic-angle graphene superlattices},\ }\href@noop {} {\bibfield  {journal} {\bibinfo  {journal} {Nature}\ }\textbf {\bibinfo {volume} {556}},\ \bibinfo {pages} {43} (\bibinfo {year} {2018}{\natexlab{a}})}\BibitemShut {NoStop}%
\bibitem [{\citenamefont {Lu}\ \emph {et~al.}(2019)\citenamefont {Lu}, \citenamefont {Stepanov}, \citenamefont {Yang}, \citenamefont {Xie}, \citenamefont {Aamir}, \citenamefont {Das}, \citenamefont {Urgell}, \citenamefont {Watanabe}, \citenamefont {Taniguchi}, \citenamefont {Zhang} \emph {et~al.}}]{lu2019superconductors}%
  \BibitemOpen
  \bibfield  {author} {\bibinfo {author} {\bibfnamefont {X.}~\bibnamefont {Lu}}, \bibinfo {author} {\bibfnamefont {P.}~\bibnamefont {Stepanov}}, \bibinfo {author} {\bibfnamefont {W.}~\bibnamefont {Yang}}, \bibinfo {author} {\bibfnamefont {M.}~\bibnamefont {Xie}}, \bibinfo {author} {\bibfnamefont {M.~A.}\ \bibnamefont {Aamir}}, \bibinfo {author} {\bibfnamefont {I.}~\bibnamefont {Das}}, \bibinfo {author} {\bibfnamefont {C.}~\bibnamefont {Urgell}}, \bibinfo {author} {\bibfnamefont {K.}~\bibnamefont {Watanabe}}, \bibinfo {author} {\bibfnamefont {T.}~\bibnamefont {Taniguchi}}, \bibinfo {author} {\bibfnamefont {G.}~\bibnamefont {Zhang}}, \emph {et~al.},\ }\bibfield  {title} {\bibinfo {title} {Superconductors, orbital magnets and correlated states in magic-angle bilayer graphene},\ }\href@noop {} {\bibfield  {journal} {\bibinfo  {journal} {Nature}\ }\textbf {\bibinfo {volume} {574}},\ \bibinfo {pages} {653} (\bibinfo {year} {2019})}\BibitemShut {NoStop}%
\bibitem [{\citenamefont {Yankowitz}\ \emph {et~al.}(2019)\citenamefont {Yankowitz}, \citenamefont {Chen}, \citenamefont {Polshyn}, \citenamefont {Zhang}, \citenamefont {Watanabe}, \citenamefont {Taniguchi}, \citenamefont {Graf}, \citenamefont {Young},\ and\ \citenamefont {Dean}}]{yankowitz2019tuning}%
  \BibitemOpen
  \bibfield  {author} {\bibinfo {author} {\bibfnamefont {M.}~\bibnamefont {Yankowitz}}, \bibinfo {author} {\bibfnamefont {S.}~\bibnamefont {Chen}}, \bibinfo {author} {\bibfnamefont {H.}~\bibnamefont {Polshyn}}, \bibinfo {author} {\bibfnamefont {Y.}~\bibnamefont {Zhang}}, \bibinfo {author} {\bibfnamefont {K.}~\bibnamefont {Watanabe}}, \bibinfo {author} {\bibfnamefont {T.}~\bibnamefont {Taniguchi}}, \bibinfo {author} {\bibfnamefont {D.}~\bibnamefont {Graf}}, \bibinfo {author} {\bibfnamefont {A.~F.}\ \bibnamefont {Young}},\ and\ \bibinfo {author} {\bibfnamefont {C.~R.}\ \bibnamefont {Dean}},\ }\bibfield  {title} {\bibinfo {title} {Tuning superconductivity in twisted bilayer graphene},\ }\href@noop {} {\bibfield  {journal} {\bibinfo  {journal} {Science}\ }\textbf {\bibinfo {volume} {363}},\ \bibinfo {pages} {1059} (\bibinfo {year} {2019})}\BibitemShut {NoStop}%
\bibitem [{\citenamefont {Cao}\ \emph {et~al.}(2018{\natexlab{b}})\citenamefont {Cao}, \citenamefont {Fatemi}, \citenamefont {Demir}, \citenamefont {Fang}, \citenamefont {Tomarken}, \citenamefont {Luo}, \citenamefont {Sanchez-Yamagishi}, \citenamefont {Watanabe}, \citenamefont {Taniguchi}, \citenamefont {Kaxiras} \emph {et~al.}}]{cao2018correlated}%
  \BibitemOpen
  \bibfield  {author} {\bibinfo {author} {\bibfnamefont {Y.}~\bibnamefont {Cao}}, \bibinfo {author} {\bibfnamefont {V.}~\bibnamefont {Fatemi}}, \bibinfo {author} {\bibfnamefont {A.}~\bibnamefont {Demir}}, \bibinfo {author} {\bibfnamefont {S.}~\bibnamefont {Fang}}, \bibinfo {author} {\bibfnamefont {S.~L.}\ \bibnamefont {Tomarken}}, \bibinfo {author} {\bibfnamefont {J.~Y.}\ \bibnamefont {Luo}}, \bibinfo {author} {\bibfnamefont {J.~D.}\ \bibnamefont {Sanchez-Yamagishi}}, \bibinfo {author} {\bibfnamefont {K.}~\bibnamefont {Watanabe}}, \bibinfo {author} {\bibfnamefont {T.}~\bibnamefont {Taniguchi}}, \bibinfo {author} {\bibfnamefont {E.}~\bibnamefont {Kaxiras}}, \emph {et~al.},\ }\bibfield  {title} {\bibinfo {title} {Correlated insulator behaviour at half-filling in magic-angle graphene superlattices},\ }\href@noop {} {\bibfield  {journal} {\bibinfo  {journal} {Nature}\ }\textbf {\bibinfo {volume} {556}},\ \bibinfo {pages} {80} (\bibinfo {year} {2018}{\natexlab{b}})}\BibitemShut {NoStop}%
\bibitem [{\citenamefont {Saito}\ \emph {et~al.}(2020)\citenamefont {Saito}, \citenamefont {Ge}, \citenamefont {Watanabe}, \citenamefont {Taniguchi},\ and\ \citenamefont {Young}}]{saito2020independent}%
  \BibitemOpen
  \bibfield  {author} {\bibinfo {author} {\bibfnamefont {Y.}~\bibnamefont {Saito}}, \bibinfo {author} {\bibfnamefont {J.}~\bibnamefont {Ge}}, \bibinfo {author} {\bibfnamefont {K.}~\bibnamefont {Watanabe}}, \bibinfo {author} {\bibfnamefont {T.}~\bibnamefont {Taniguchi}},\ and\ \bibinfo {author} {\bibfnamefont {A.~F.}\ \bibnamefont {Young}},\ }\bibfield  {title} {\bibinfo {title} {Independent superconductors and correlated insulators in twisted bilayer graphene},\ }\href@noop {} {\bibfield  {journal} {\bibinfo  {journal} {Nature Physics}\ }\textbf {\bibinfo {volume} {16}},\ \bibinfo {pages} {926} (\bibinfo {year} {2020})}\BibitemShut {NoStop}%
\bibitem [{\citenamefont {Serlin}\ \emph {et~al.}(2020)\citenamefont {Serlin}, \citenamefont {Tschirhart}, \citenamefont {Polshyn}, \citenamefont {Zhang}, \citenamefont {Zhu}, \citenamefont {Watanabe}, \citenamefont {Taniguchi}, \citenamefont {Balents},\ and\ \citenamefont {Young}}]{serlin2020intrinsic}%
  \BibitemOpen
  \bibfield  {author} {\bibinfo {author} {\bibfnamefont {M.}~\bibnamefont {Serlin}}, \bibinfo {author} {\bibfnamefont {C.}~\bibnamefont {Tschirhart}}, \bibinfo {author} {\bibfnamefont {H.}~\bibnamefont {Polshyn}}, \bibinfo {author} {\bibfnamefont {Y.}~\bibnamefont {Zhang}}, \bibinfo {author} {\bibfnamefont {J.}~\bibnamefont {Zhu}}, \bibinfo {author} {\bibfnamefont {K.}~\bibnamefont {Watanabe}}, \bibinfo {author} {\bibfnamefont {T.}~\bibnamefont {Taniguchi}}, \bibinfo {author} {\bibfnamefont {L.}~\bibnamefont {Balents}},\ and\ \bibinfo {author} {\bibfnamefont {A.}~\bibnamefont {Young}},\ }\bibfield  {title} {\bibinfo {title} {Intrinsic quantized anomalous hall effect in a moir{\'e} heterostructure},\ }\href@noop {} {\bibfield  {journal} {\bibinfo  {journal} {Science}\ }\textbf {\bibinfo {volume} {367}},\ \bibinfo {pages} {900} (\bibinfo {year} {2020})}\BibitemShut {NoStop}%
\bibitem [{\citenamefont {Li}\ \emph {et~al.}(2021)\citenamefont {Li}, \citenamefont {Jiang}, \citenamefont {Shen}, \citenamefont {Zhang}, \citenamefont {Li}, \citenamefont {Tao}, \citenamefont {Devakul}, \citenamefont {Watanabe}, \citenamefont {Taniguchi}, \citenamefont {Fu} \emph {et~al.}}]{li2021quantum}%
  \BibitemOpen
  \bibfield  {author} {\bibinfo {author} {\bibfnamefont {T.}~\bibnamefont {Li}}, \bibinfo {author} {\bibfnamefont {S.}~\bibnamefont {Jiang}}, \bibinfo {author} {\bibfnamefont {B.}~\bibnamefont {Shen}}, \bibinfo {author} {\bibfnamefont {Y.}~\bibnamefont {Zhang}}, \bibinfo {author} {\bibfnamefont {L.}~\bibnamefont {Li}}, \bibinfo {author} {\bibfnamefont {Z.}~\bibnamefont {Tao}}, \bibinfo {author} {\bibfnamefont {T.}~\bibnamefont {Devakul}}, \bibinfo {author} {\bibfnamefont {K.}~\bibnamefont {Watanabe}}, \bibinfo {author} {\bibfnamefont {T.}~\bibnamefont {Taniguchi}}, \bibinfo {author} {\bibfnamefont {L.}~\bibnamefont {Fu}}, \emph {et~al.},\ }\bibfield  {title} {\bibinfo {title} {Quantum anomalous hall effect from intertwined moir{\'e} bands},\ }\href@noop {} {\bibfield  {journal} {\bibinfo  {journal} {Nature}\ }\textbf {\bibinfo {volume} {600}},\ \bibinfo {pages} {641} (\bibinfo {year} {2021})}\BibitemShut {NoStop}%
\bibitem [{\citenamefont {Cai}\ \emph {et~al.}(2023)\citenamefont {Cai}, \citenamefont {Anderson}, \citenamefont {Wang}, \citenamefont {Zhang}, \citenamefont {Liu}, \citenamefont {Holtzmann}, \citenamefont {Zhang}, \citenamefont {Fan}, \citenamefont {Taniguchi}, \citenamefont {Watanabe} \emph {et~al.}}]{cai2023signatures}%
  \BibitemOpen
  \bibfield  {author} {\bibinfo {author} {\bibfnamefont {J.}~\bibnamefont {Cai}}, \bibinfo {author} {\bibfnamefont {E.}~\bibnamefont {Anderson}}, \bibinfo {author} {\bibfnamefont {C.}~\bibnamefont {Wang}}, \bibinfo {author} {\bibfnamefont {X.}~\bibnamefont {Zhang}}, \bibinfo {author} {\bibfnamefont {X.}~\bibnamefont {Liu}}, \bibinfo {author} {\bibfnamefont {W.}~\bibnamefont {Holtzmann}}, \bibinfo {author} {\bibfnamefont {Y.}~\bibnamefont {Zhang}}, \bibinfo {author} {\bibfnamefont {F.}~\bibnamefont {Fan}}, \bibinfo {author} {\bibfnamefont {T.}~\bibnamefont {Taniguchi}}, \bibinfo {author} {\bibfnamefont {K.}~\bibnamefont {Watanabe}}, \emph {et~al.},\ }\bibfield  {title} {\bibinfo {title} {Signatures of fractional quantum anomalous hall states in twisted mote2},\ }\href@noop {} {\bibfield  {journal} {\bibinfo  {journal} {Nature}\ }\textbf {\bibinfo {volume} {622}},\ \bibinfo {pages} {63} (\bibinfo {year} {2023})}\BibitemShut {NoStop}%
\bibitem [{\citenamefont {Park}\ \emph {et~al.}(2023)\citenamefont {Park}, \citenamefont {Cai}, \citenamefont {Anderson}, \citenamefont {Zhang}, \citenamefont {Zhu}, \citenamefont {Liu}, \citenamefont {Wang}, \citenamefont {Holtzmann}, \citenamefont {Hu}, \citenamefont {Liu} \emph {et~al.}}]{park2023observation}%
  \BibitemOpen
  \bibfield  {author} {\bibinfo {author} {\bibfnamefont {H.}~\bibnamefont {Park}}, \bibinfo {author} {\bibfnamefont {J.}~\bibnamefont {Cai}}, \bibinfo {author} {\bibfnamefont {E.}~\bibnamefont {Anderson}}, \bibinfo {author} {\bibfnamefont {Y.}~\bibnamefont {Zhang}}, \bibinfo {author} {\bibfnamefont {J.}~\bibnamefont {Zhu}}, \bibinfo {author} {\bibfnamefont {X.}~\bibnamefont {Liu}}, \bibinfo {author} {\bibfnamefont {C.}~\bibnamefont {Wang}}, \bibinfo {author} {\bibfnamefont {W.}~\bibnamefont {Holtzmann}}, \bibinfo {author} {\bibfnamefont {C.}~\bibnamefont {Hu}}, \bibinfo {author} {\bibfnamefont {Z.}~\bibnamefont {Liu}}, \emph {et~al.},\ }\bibfield  {title} {\bibinfo {title} {Observation of fractionally quantized anomalous hall effect},\ }\href@noop {} {\bibfield  {journal} {\bibinfo  {journal} {Nature}\ }\textbf {\bibinfo {volume} {622}},\ \bibinfo {pages} {74} (\bibinfo {year} {2023})}\BibitemShut {NoStop}%
\bibitem [{\citenamefont {Lu}\ \emph {et~al.}(2024{\natexlab{a}})\citenamefont {Lu}, \citenamefont {Han}, \citenamefont {Yao}, \citenamefont {Reddy}, \citenamefont {Yang}, \citenamefont {Seo}, \citenamefont {Watanabe}, \citenamefont {Taniguchi}, \citenamefont {Fu},\ and\ \citenamefont {Ju}}]{lu2024fractional}%
  \BibitemOpen
  \bibfield  {author} {\bibinfo {author} {\bibfnamefont {Z.}~\bibnamefont {Lu}}, \bibinfo {author} {\bibfnamefont {T.}~\bibnamefont {Han}}, \bibinfo {author} {\bibfnamefont {Y.}~\bibnamefont {Yao}}, \bibinfo {author} {\bibfnamefont {A.~P.}\ \bibnamefont {Reddy}}, \bibinfo {author} {\bibfnamefont {J.}~\bibnamefont {Yang}}, \bibinfo {author} {\bibfnamefont {J.}~\bibnamefont {Seo}}, \bibinfo {author} {\bibfnamefont {K.}~\bibnamefont {Watanabe}}, \bibinfo {author} {\bibfnamefont {T.}~\bibnamefont {Taniguchi}}, \bibinfo {author} {\bibfnamefont {L.}~\bibnamefont {Fu}},\ and\ \bibinfo {author} {\bibfnamefont {L.}~\bibnamefont {Ju}},\ }\bibfield  {title} {\bibinfo {title} {Fractional quantum anomalous hall effect in multilayer graphene},\ }\href@noop {} {\bibfield  {journal} {\bibinfo  {journal} {Nature}\ }\textbf {\bibinfo {volume} {626}},\ \bibinfo {pages} {759} (\bibinfo {year} {2024}{\natexlab{a}})}\BibitemShut {NoStop}%
\bibitem [{\citenamefont {Xu}\ \emph {et~al.}(2023)\citenamefont {Xu}, \citenamefont {Sun}, \citenamefont {Jia}, \citenamefont {Liu}, \citenamefont {Xu}, \citenamefont {Li}, \citenamefont {Gu}, \citenamefont {Watanabe}, \citenamefont {Taniguchi}, \citenamefont {Tong} \emph {et~al.}}]{xu2023observation}%
  \BibitemOpen
  \bibfield  {author} {\bibinfo {author} {\bibfnamefont {F.}~\bibnamefont {Xu}}, \bibinfo {author} {\bibfnamefont {Z.}~\bibnamefont {Sun}}, \bibinfo {author} {\bibfnamefont {T.}~\bibnamefont {Jia}}, \bibinfo {author} {\bibfnamefont {C.}~\bibnamefont {Liu}}, \bibinfo {author} {\bibfnamefont {C.}~\bibnamefont {Xu}}, \bibinfo {author} {\bibfnamefont {C.}~\bibnamefont {Li}}, \bibinfo {author} {\bibfnamefont {Y.}~\bibnamefont {Gu}}, \bibinfo {author} {\bibfnamefont {K.}~\bibnamefont {Watanabe}}, \bibinfo {author} {\bibfnamefont {T.}~\bibnamefont {Taniguchi}}, \bibinfo {author} {\bibfnamefont {B.}~\bibnamefont {Tong}}, \emph {et~al.},\ }\bibfield  {title} {\bibinfo {title} {Observation of integer and fractional quantum anomalous hall effects in twisted bilayer mote 2},\ }\href@noop {} {\bibfield  {journal} {\bibinfo  {journal} {Physical Review X}\ }\textbf {\bibinfo {volume} {13}},\ \bibinfo {pages} {031037} (\bibinfo {year} {2023})}\BibitemShut {NoStop}%
\bibitem [{\citenamefont {Zhang}\ \emph {et~al.}(2024)\citenamefont {Zhang}, \citenamefont {Yang}, \citenamefont {Xie}, \citenamefont {Feng}, \citenamefont {Zhang}, \citenamefont {Watanabe}, \citenamefont {Taniguchi}, \citenamefont {Yang}, \citenamefont {Dai}, \citenamefont {Liu}, \citenamefont {Liu}, \citenamefont {Liu}, \citenamefont {Song}, \citenamefont {Liu},\ and\ \citenamefont {Lu}}]{zhang_tbg_qah_2024}%
  \BibitemOpen
  \bibfield  {author} {\bibinfo {author} {\bibfnamefont {Z.}~\bibnamefont {Zhang}}, \bibinfo {author} {\bibfnamefont {J.}~\bibnamefont {Yang}}, \bibinfo {author} {\bibfnamefont {B.}~\bibnamefont {Xie}}, \bibinfo {author} {\bibfnamefont {Z.}~\bibnamefont {Feng}}, \bibinfo {author} {\bibfnamefont {S.}~\bibnamefont {Zhang}}, \bibinfo {author} {\bibfnamefont {K.}~\bibnamefont {Watanabe}}, \bibinfo {author} {\bibfnamefont {T.}~\bibnamefont {Taniguchi}}, \bibinfo {author} {\bibfnamefont {X.}~\bibnamefont {Yang}}, \bibinfo {author} {\bibfnamefont {Q.}~\bibnamefont {Dai}}, \bibinfo {author} {\bibfnamefont {T.}~\bibnamefont {Liu}}, \bibinfo {author} {\bibfnamefont {D.}~\bibnamefont {Liu}}, \bibinfo {author} {\bibfnamefont {K.}~\bibnamefont {Liu}}, \bibinfo {author} {\bibfnamefont {Z.}~\bibnamefont {Song}}, \bibinfo {author} {\bibfnamefont {J.}~\bibnamefont {Liu}},\ and\ \bibinfo {author} {\bibfnamefont {X.}~\bibnamefont {Lu}},\ }\href@noop {} {\bibinfo {title} {Commensurate and incommensurate chern insulators in
  magic-angle bilayer graphene}} (\bibinfo {year} {2024})\BibitemShut {NoStop}%
\bibitem [{\citenamefont {Wan}\ \emph {et~al.}(2023)\citenamefont {Wan}, \citenamefont {Sarkar}, \citenamefont {Sun},\ and\ \citenamefont {Lin}}]{wan_nearly_2023}%
  \BibitemOpen
  \bibfield  {author} {\bibinfo {author} {\bibfnamefont {X.}~\bibnamefont {Wan}}, \bibinfo {author} {\bibfnamefont {S.}~\bibnamefont {Sarkar}}, \bibinfo {author} {\bibfnamefont {K.}~\bibnamefont {Sun}},\ and\ \bibinfo {author} {\bibfnamefont {S.-Z.}\ \bibnamefont {Lin}},\ }\bibfield  {title} {\bibinfo {title} {Nearly flat {Chern} band in periodically strained monolayer and bilayer graphene},\ }\href@noop {} {\bibfield  {journal} {\bibinfo  {journal} {Phys. Rev. B}\ }\textbf {\bibinfo {volume} {108}},\ \bibinfo {pages} {125129} (\bibinfo {year} {2023})}\BibitemShut {NoStop}%
\bibitem [{\citenamefont {Gao}\ \emph {et~al.}(2023)\citenamefont {Gao}, \citenamefont {Dong}, \citenamefont {Ledwith}, \citenamefont {Parker},\ and\ \citenamefont {Khalaf}}]{gao_untwisting_2023}%
  \BibitemOpen
  \bibfield  {author} {\bibinfo {author} {\bibfnamefont {Q.}~\bibnamefont {Gao}}, \bibinfo {author} {\bibfnamefont {J.}~\bibnamefont {Dong}}, \bibinfo {author} {\bibfnamefont {P.}~\bibnamefont {Ledwith}}, \bibinfo {author} {\bibfnamefont {D.}~\bibnamefont {Parker}},\ and\ \bibinfo {author} {\bibfnamefont {E.}~\bibnamefont {Khalaf}},\ }\bibfield  {title} {\bibinfo {title} {Untwisting {Moir}{\textbackslash}'e {Physics}: {Almost} {Ideal} {Bands} and {Fractional} {Chern} {Insulators} in {Periodically} {Strained} {Monolayer} {Graphene}},\ }\href@noop {} {\bibfield  {journal} {\bibinfo  {journal} {Phys. Rev. Lett.}\ }\textbf {\bibinfo {volume} {131}},\ \bibinfo {pages} {096401} (\bibinfo {year} {2023})}\BibitemShut {NoStop}%
\bibitem [{\citenamefont {Lian}\ \emph {et~al.}(2020)\citenamefont {Lian}, \citenamefont {Liu}, \citenamefont {Zhang},\ and\ \citenamefont {Wang}}]{lian_flat_2020}%
  \BibitemOpen
  \bibfield  {author} {\bibinfo {author} {\bibfnamefont {B.}~\bibnamefont {Lian}}, \bibinfo {author} {\bibfnamefont {Z.}~\bibnamefont {Liu}}, \bibinfo {author} {\bibfnamefont {Y.}~\bibnamefont {Zhang}},\ and\ \bibinfo {author} {\bibfnamefont {J.}~\bibnamefont {Wang}},\ }\bibfield  {title} {\bibinfo {title} {Flat {Chern} {Band} from {Twisted} {Bilayer} \ce{MnBi2Te4}},\ }\href@noop {} {\bibfield  {journal} {\bibinfo  {journal} {Phys. Rev. Lett.}\ }\textbf {\bibinfo {volume} {124}},\ \bibinfo {pages} {126402} (\bibinfo {year} {2020})}\BibitemShut {NoStop}%
\bibitem [{\citenamefont {Lu}\ \emph {et~al.}(2024{\natexlab{b}})\citenamefont {Lu}, \citenamefont {Xie}, \citenamefont {Yang}, \citenamefont {Zhang}, \citenamefont {Kong}, \citenamefont {Li}, \citenamefont {Ding}, \citenamefont {Wang},\ and\ \citenamefont {Liu}}]{lu_magic_2024}%
  \BibitemOpen
  \bibfield  {author} {\bibinfo {author} {\bibfnamefont {X.}~\bibnamefont {Lu}}, \bibinfo {author} {\bibfnamefont {B.}~\bibnamefont {Xie}}, \bibinfo {author} {\bibfnamefont {Y.}~\bibnamefont {Yang}}, \bibinfo {author} {\bibfnamefont {Y.}~\bibnamefont {Zhang}}, \bibinfo {author} {\bibfnamefont {X.}~\bibnamefont {Kong}}, \bibinfo {author} {\bibfnamefont {J.}~\bibnamefont {Li}}, \bibinfo {author} {\bibfnamefont {F.}~\bibnamefont {Ding}}, \bibinfo {author} {\bibfnamefont {Z.-J.}\ \bibnamefont {Wang}},\ and\ \bibinfo {author} {\bibfnamefont {J.}~\bibnamefont {Liu}},\ }\bibfield  {title} {\bibinfo {title} {Magic {Momenta} and {Three}-{Dimensional} {Landau} {Levels} from a {Three}-{Dimensional} {Graphite} {Moir\'e} {Superlattice}},\ }\href@noop {} {\bibfield  {journal} {\bibinfo  {journal} {Phys. Rev. Lett.}\ }\textbf {\bibinfo {volume} {132}},\ \bibinfo {pages} {056601} (\bibinfo {year} {2024}{\natexlab{b}})}\BibitemShut {NoStop}%
\bibitem [{\citenamefont {Yang}\ \emph {et~al.}(2024{\natexlab{a}})\citenamefont {Yang}, \citenamefont {Liu}, \citenamefont {Schindler},\ and\ \citenamefont {Liu}}]{yang_engineering_2024}%
  \BibitemOpen
  \bibfield  {author} {\bibinfo {author} {\bibfnamefont {K.}~\bibnamefont {Yang}}, \bibinfo {author} {\bibfnamefont {Y.}~\bibnamefont {Liu}}, \bibinfo {author} {\bibfnamefont {F.}~\bibnamefont {Schindler}},\ and\ \bibinfo {author} {\bibfnamefont {C.-X.}\ \bibnamefont {Liu}},\ }\href@noop {} {\bibinfo {title} {Engineering {Miniband} {Topology} via {Band}-{Folding} in {Moir}{\textbackslash}'e {Superlattice} {Materials}}} (\bibinfo {year} {2024}{\natexlab{a}})\BibitemShut {NoStop}%
\bibitem [{\citenamefont {Ghorashi}\ \emph {et~al.}(2023)\citenamefont {Ghorashi}, \citenamefont {Dunbrack}, \citenamefont {Abouelkomsan}, \citenamefont {Sun}, \citenamefont {Du},\ and\ \citenamefont {Cano}}]{ghorashi_topological_2023}%
  \BibitemOpen
  \bibfield  {author} {\bibinfo {author} {\bibfnamefont {S.~A.~A.}\ \bibnamefont {Ghorashi}}, \bibinfo {author} {\bibfnamefont {A.}~\bibnamefont {Dunbrack}}, \bibinfo {author} {\bibfnamefont {A.}~\bibnamefont {Abouelkomsan}}, \bibinfo {author} {\bibfnamefont {J.}~\bibnamefont {Sun}}, \bibinfo {author} {\bibfnamefont {X.}~\bibnamefont {Du}},\ and\ \bibinfo {author} {\bibfnamefont {J.}~\bibnamefont {Cano}},\ }\bibfield  {title} {\bibinfo {title} {Topological and {Stacked} {Flat} {Bands} in {Bilayer} {Graphene} with a {Superlattice} {Potential}},\ }\href@noop {} {\bibfield  {journal} {\bibinfo  {journal} {Phys. Rev. Lett.}\ }\textbf {\bibinfo {volume} {130}},\ \bibinfo {pages} {196201} (\bibinfo {year} {2023})}\BibitemShut {NoStop}%
\bibitem [{\citenamefont {Wang}\ \emph {et~al.}(2024{\natexlab{a}})\citenamefont {Wang}, \citenamefont {Zhan}, \citenamefont {Fan}, \citenamefont {Li}, \citenamefont {Pantaleón}, \citenamefont {Ye}, \citenamefont {He}, \citenamefont {Wei}, \citenamefont {Li}, \citenamefont {Guinea}, \citenamefont {Yuan},\ and\ \citenamefont {Zeng}}]{wang_dispersion-selective_2024}%
  \BibitemOpen
  \bibfield  {author} {\bibinfo {author} {\bibfnamefont {S.}~\bibnamefont {Wang}}, \bibinfo {author} {\bibfnamefont {Z.}~\bibnamefont {Zhan}}, \bibinfo {author} {\bibfnamefont {X.}~\bibnamefont {Fan}}, \bibinfo {author} {\bibfnamefont {Y.}~\bibnamefont {Li}}, \bibinfo {author} {\bibfnamefont {P.~A.}\ \bibnamefont {Pantaleón}}, \bibinfo {author} {\bibfnamefont {C.}~\bibnamefont {Ye}}, \bibinfo {author} {\bibfnamefont {Z.}~\bibnamefont {He}}, \bibinfo {author} {\bibfnamefont {L.}~\bibnamefont {Wei}}, \bibinfo {author} {\bibfnamefont {L.}~\bibnamefont {Li}}, \bibinfo {author} {\bibfnamefont {F.}~\bibnamefont {Guinea}}, \bibinfo {author} {\bibfnamefont {S.}~\bibnamefont {Yuan}},\ and\ \bibinfo {author} {\bibfnamefont {C.}~\bibnamefont {Zeng}},\ }\bibfield  {title} {\bibinfo {title} {Dispersion-{Selective} {Band} {Engineering} in an {Artificial} {Kagome} {Superlattice}},\ }\href@noop {} {\bibfield  {journal} {\bibinfo  {journal} {Phys. Rev. Lett.}\ }\textbf {\bibinfo {volume} {133}},\ \bibinfo {pages} {066302}
  (\bibinfo {year} {2024}{\natexlab{a}})}\BibitemShut {NoStop}%
\bibitem [{\citenamefont {Wang}\ \emph {et~al.}(2024{\natexlab{b}})\citenamefont {Wang}, \citenamefont {Fan},\ and\ \citenamefont {Zhang}}]{wang_electrically_2024}%
  \BibitemOpen
  \bibfield  {author} {\bibinfo {author} {\bibfnamefont {H.}~\bibnamefont {Wang}}, \bibinfo {author} {\bibfnamefont {Y.}~\bibnamefont {Fan}},\ and\ \bibinfo {author} {\bibfnamefont {H.}~\bibnamefont {Zhang}},\ }\bibfield  {title} {\bibinfo {title} {Electrically tunable high-{Chern}-number quasiflat bands in twisted antiferromagnetic topological insulators},\ }\href@noop {} {\bibfield  {journal} {\bibinfo  {journal} {Phys. Rev. B}\ }\textbf {\bibinfo {volume} {110}},\ \bibinfo {pages} {085135} (\bibinfo {year} {2024}{\natexlab{b}})}\BibitemShut {NoStop}%
\bibitem [{\citenamefont {Anderson}\ \emph {et~al.}(2023)\citenamefont {Anderson}, \citenamefont {Fan}, \citenamefont {Cai}, \citenamefont {Holtzmann}, \citenamefont {Taniguchi}, \citenamefont {Watanabe}, \citenamefont {Xiao}, \citenamefont {Yao},\ and\ \citenamefont {Xu}}]{anderson_programming_2023}%
  \BibitemOpen
  \bibfield  {author} {\bibinfo {author} {\bibfnamefont {E.}~\bibnamefont {Anderson}}, \bibinfo {author} {\bibfnamefont {F.-R.}\ \bibnamefont {Fan}}, \bibinfo {author} {\bibfnamefont {J.}~\bibnamefont {Cai}}, \bibinfo {author} {\bibfnamefont {W.}~\bibnamefont {Holtzmann}}, \bibinfo {author} {\bibfnamefont {T.}~\bibnamefont {Taniguchi}}, \bibinfo {author} {\bibfnamefont {K.}~\bibnamefont {Watanabe}}, \bibinfo {author} {\bibfnamefont {D.}~\bibnamefont {Xiao}}, \bibinfo {author} {\bibfnamefont {W.}~\bibnamefont {Yao}},\ and\ \bibinfo {author} {\bibfnamefont {X.}~\bibnamefont {Xu}},\ }\href@noop {} {\bibinfo {title} {Programming {Correlated} {Magnetic} {States} via {Gate} {Controlled} {Moir}{\textbackslash}'e {Geometry}}} (\bibinfo {year} {2023})\BibitemShut {NoStop}%
\bibitem [{\citenamefont {Lu}\ \emph {et~al.}(2023)\citenamefont {Lu}, \citenamefont {Zhang}, \citenamefont {Wang}, \citenamefont {Gao}, \citenamefont {Yang}, \citenamefont {Guo}, \citenamefont {Gao}, \citenamefont {Ye}, \citenamefont {Han},\ and\ \citenamefont {Liu}}]{Lu_nc_2023}%
  \BibitemOpen
  \bibfield  {author} {\bibinfo {author} {\bibfnamefont {X.}~\bibnamefont {Lu}}, \bibinfo {author} {\bibfnamefont {S.}~\bibnamefont {Zhang}}, \bibinfo {author} {\bibfnamefont {Y.}~\bibnamefont {Wang}}, \bibinfo {author} {\bibfnamefont {X.}~\bibnamefont {Gao}}, \bibinfo {author} {\bibfnamefont {K.}~\bibnamefont {Yang}}, \bibinfo {author} {\bibfnamefont {Z.}~\bibnamefont {Guo}}, \bibinfo {author} {\bibfnamefont {Y.}~\bibnamefont {Gao}}, \bibinfo {author} {\bibfnamefont {Y.}~\bibnamefont {Ye}}, \bibinfo {author} {\bibfnamefont {Z.}~\bibnamefont {Han}},\ and\ \bibinfo {author} {\bibfnamefont {J.}~\bibnamefont {Liu}},\ }\bibfield  {title} {\bibinfo {title} {Synergistic correlated states and nontrivial topology in coupled graphene-insulator heterostructures},\ }\href@noop {} {\bibfield  {journal} {\bibinfo  {journal} {Nature Communications}\ }\textbf {\bibinfo {volume} {14}} (\bibinfo {year} {2023})}\BibitemShut {NoStop}%
\bibitem [{\citenamefont {Kim}\ \emph {et~al.}(2023)\citenamefont {Kim}, \citenamefont {Dominguez}, \citenamefont {Mayorga-Luna}, \citenamefont {Ye}, \citenamefont {Embley}, \citenamefont {Tan}, \citenamefont {Ni}, \citenamefont {Liu}, \citenamefont {Ford}, \citenamefont {Gao}, \citenamefont {Arash}, \citenamefont {Watanabe}, \citenamefont {Taniguchi}, \citenamefont {Kim}, \citenamefont {Shih}, \citenamefont {Lai}, \citenamefont {Yao}, \citenamefont {Yang}, \citenamefont {Li},\ and\ \citenamefont {Miyahara}}]{Kim2023}%
  \BibitemOpen
  \bibfield  {author} {\bibinfo {author} {\bibfnamefont {D.~S.}\ \bibnamefont {Kim}}, \bibinfo {author} {\bibfnamefont {R.~C.}\ \bibnamefont {Dominguez}}, \bibinfo {author} {\bibfnamefont {R.}~\bibnamefont {Mayorga-Luna}}, \bibinfo {author} {\bibfnamefont {D.}~\bibnamefont {Ye}}, \bibinfo {author} {\bibfnamefont {J.}~\bibnamefont {Embley}}, \bibinfo {author} {\bibfnamefont {T.}~\bibnamefont {Tan}}, \bibinfo {author} {\bibfnamefont {Y.}~\bibnamefont {Ni}}, \bibinfo {author} {\bibfnamefont {Z.}~\bibnamefont {Liu}}, \bibinfo {author} {\bibfnamefont {M.}~\bibnamefont {Ford}}, \bibinfo {author} {\bibfnamefont {F.~Y.}\ \bibnamefont {Gao}}, \bibinfo {author} {\bibfnamefont {S.}~\bibnamefont {Arash}}, \bibinfo {author} {\bibfnamefont {K.}~\bibnamefont {Watanabe}}, \bibinfo {author} {\bibfnamefont {T.}~\bibnamefont {Taniguchi}}, \bibinfo {author} {\bibfnamefont {S.}~\bibnamefont {Kim}}, \bibinfo {author} {\bibfnamefont {C.-K.}\ \bibnamefont {Shih}}, \bibinfo {author} {\bibfnamefont {K.}~\bibnamefont {Lai}}, \bibinfo
  {author} {\bibfnamefont {W.}~\bibnamefont {Yao}}, \bibinfo {author} {\bibfnamefont {L.}~\bibnamefont {Yang}}, \bibinfo {author} {\bibfnamefont {X.}~\bibnamefont {Li}},\ and\ \bibinfo {author} {\bibfnamefont {Y.}~\bibnamefont {Miyahara}},\ }\bibfield  {title} {\bibinfo {title} {Electrostatic moiré potential from twisted hexagonal boron nitride layers},\ }\href@noop {} {\bibfield  {journal} {\bibinfo  {journal} {Nature Materials}\ }\textbf {\bibinfo {volume} {23}},\ \bibinfo {pages} {65–70} (\bibinfo {year} {2023})}\BibitemShut {NoStop}%
\bibitem [{\citenamefont {Rashidi}\ \emph {et~al.}(2024)\citenamefont {Rashidi}, \citenamefont {Ahadi}, \citenamefont {Munyan}, \citenamefont {Mitchell},\ and\ \citenamefont {Stemmer}}]{rashidi_tuning_2024}%
  \BibitemOpen
  \bibfield  {author} {\bibinfo {author} {\bibfnamefont {A.}~\bibnamefont {Rashidi}}, \bibinfo {author} {\bibfnamefont {S.}~\bibnamefont {Ahadi}}, \bibinfo {author} {\bibfnamefont {S.}~\bibnamefont {Munyan}}, \bibinfo {author} {\bibfnamefont {W.~J.}\ \bibnamefont {Mitchell}},\ and\ \bibinfo {author} {\bibfnamefont {S.}~\bibnamefont {Stemmer}},\ }\bibfield  {title} {\bibinfo {title} {Tuning {Displacement} {Fields} in a {Two}-{Dimensional} {Topological} {Insulator} {Using} {Nanopatterned} {Gates}},\ }\href@noop {} {\bibfield  {journal} {\bibinfo  {journal} {Nano Lett.}\ }\textbf {\bibinfo {volume} {24}},\ \bibinfo {pages} {7366} (\bibinfo {year} {2024})}\BibitemShut {NoStop}%
\bibitem [{\citenamefont {Sun}\ \emph {et~al.}(2023)\citenamefont {Sun}, \citenamefont {Ghorashi}, \citenamefont {Watanabe}, \citenamefont {Taniguchi}, \citenamefont {Camino}, \citenamefont {Cano},\ and\ \citenamefont {Du}}]{sun_signature_2023}%
  \BibitemOpen
  \bibfield  {author} {\bibinfo {author} {\bibfnamefont {J.}~\bibnamefont {Sun}}, \bibinfo {author} {\bibfnamefont {S.~A.~A.}\ \bibnamefont {Ghorashi}}, \bibinfo {author} {\bibfnamefont {K.}~\bibnamefont {Watanabe}}, \bibinfo {author} {\bibfnamefont {T.}~\bibnamefont {Taniguchi}}, \bibinfo {author} {\bibfnamefont {F.}~\bibnamefont {Camino}}, \bibinfo {author} {\bibfnamefont {J.}~\bibnamefont {Cano}},\ and\ \bibinfo {author} {\bibfnamefont {X.}~\bibnamefont {Du}},\ }\href@noop {} {\bibinfo {title} {Signature of {Correlated} {Insulator} in {Electric} {Field} {Controlled} {Superlattice}}} (\bibinfo {year} {2023})\BibitemShut {NoStop}%
\bibitem [{\citenamefont {Su}\ \emph {et~al.}(2022)\citenamefont {Su}, \citenamefont {Li}, \citenamefont {Zhang}, \citenamefont {Sun},\ and\ \citenamefont {Lin}}]{su_massive_2022}%
  \BibitemOpen
  \bibfield  {author} {\bibinfo {author} {\bibfnamefont {Y.}~\bibnamefont {Su}}, \bibinfo {author} {\bibfnamefont {H.}~\bibnamefont {Li}}, \bibinfo {author} {\bibfnamefont {C.}~\bibnamefont {Zhang}}, \bibinfo {author} {\bibfnamefont {K.}~\bibnamefont {Sun}},\ and\ \bibinfo {author} {\bibfnamefont {S.-Z.}\ \bibnamefont {Lin}},\ }\bibfield  {title} {\bibinfo {title} {Massive {Dirac} fermions in moir{\textbackslash}'e superlattices: {A} route towards topological flat minibands and correlated topological insulators},\ }\href@noop {} {\bibfield  {journal} {\bibinfo  {journal} {Phys. Rev. Res.}\ }\textbf {\bibinfo {volume} {4}},\ \bibinfo {pages} {L032024} (\bibinfo {year} {2022})}\BibitemShut {NoStop}%
\bibitem [{\citenamefont {Suri}\ \emph {et~al.}(2023)\citenamefont {Suri}, \citenamefont {Wang}, \citenamefont {Hunt},\ and\ \citenamefont {Xiao}}]{suri_superlattice_2023}%
  \BibitemOpen
  \bibfield  {author} {\bibinfo {author} {\bibfnamefont {N.}~\bibnamefont {Suri}}, \bibinfo {author} {\bibfnamefont {C.}~\bibnamefont {Wang}}, \bibinfo {author} {\bibfnamefont {B.~M.}\ \bibnamefont {Hunt}},\ and\ \bibinfo {author} {\bibfnamefont {D.}~\bibnamefont {Xiao}},\ }\href@noop {} {\bibinfo {title} {Superlattice {Engineering} of {Topology} in {Massive} {Dirac} {Fermions}}} (\bibinfo {year} {2023})\BibitemShut {NoStop}%
\bibitem [{\citenamefont {Cano}\ \emph {et~al.}(2021)\citenamefont {Cano}, \citenamefont {Fang}, \citenamefont {Pixley},\ and\ \citenamefont {Wilson}}]{cano_moire_2021}%
  \BibitemOpen
  \bibfield  {author} {\bibinfo {author} {\bibfnamefont {J.}~\bibnamefont {Cano}}, \bibinfo {author} {\bibfnamefont {S.}~\bibnamefont {Fang}}, \bibinfo {author} {\bibfnamefont {J.~H.}\ \bibnamefont {Pixley}},\ and\ \bibinfo {author} {\bibfnamefont {J.~H.}\ \bibnamefont {Wilson}},\ }\bibfield  {title} {\bibinfo {title} {Moir{\textbackslash}'e superlattice on the surface of a topological insulator},\ }\href@noop {} {\bibfield  {journal} {\bibinfo  {journal} {Phys. Rev. B}\ }\textbf {\bibinfo {volume} {103}},\ \bibinfo {pages} {155157} (\bibinfo {year} {2021})}\BibitemShut {NoStop}%
\bibitem [{\citenamefont {Wang}\ \emph {et~al.}(2021)\citenamefont {Wang}, \citenamefont {Yuan},\ and\ \citenamefont {Fu}}]{wang_moire_2021}%
  \BibitemOpen
  \bibfield  {author} {\bibinfo {author} {\bibfnamefont {T.}~\bibnamefont {Wang}}, \bibinfo {author} {\bibfnamefont {N.~F.~Q.}\ \bibnamefont {Yuan}},\ and\ \bibinfo {author} {\bibfnamefont {L.}~\bibnamefont {Fu}},\ }\bibfield  {title} {\bibinfo {title} {Moir{\textbackslash}'e {Surface} {States} and {Enhanced} {Superconductivity} in {Topological} {Insulators}},\ }\href@noop {} {\bibfield  {journal} {\bibinfo  {journal} {Phys. Rev. X}\ }\textbf {\bibinfo {volume} {11}},\ \bibinfo {pages} {021024} (\bibinfo {year} {2021})}\BibitemShut {NoStop}%
\bibitem [{\citenamefont {Zeng}\ \emph {et~al.}(2024{\natexlab{a}})\citenamefont {Zeng}, \citenamefont {Wolf}, \citenamefont {Huang}, \citenamefont {Wei}, \citenamefont {Ghorashi}, \citenamefont {MacDonald},\ and\ \citenamefont {Cano}}]{zeng_gate-tunable_2024}%
  \BibitemOpen
  \bibfield  {author} {\bibinfo {author} {\bibfnamefont {Y.}~\bibnamefont {Zeng}}, \bibinfo {author} {\bibfnamefont {T.~M.~R.}\ \bibnamefont {Wolf}}, \bibinfo {author} {\bibfnamefont {C.}~\bibnamefont {Huang}}, \bibinfo {author} {\bibfnamefont {N.}~\bibnamefont {Wei}}, \bibinfo {author} {\bibfnamefont {S.~A.~A.}\ \bibnamefont {Ghorashi}}, \bibinfo {author} {\bibfnamefont {A.~H.}\ \bibnamefont {MacDonald}},\ and\ \bibinfo {author} {\bibfnamefont {J.}~\bibnamefont {Cano}},\ }\bibfield  {title} {\bibinfo {title} {Gate-tunable topological phases in superlattice modulated bilayer graphene},\ }\href@noop {} {\bibfield  {journal} {\bibinfo  {journal} {Phys. Rev. B}\ }\textbf {\bibinfo {volume} {109}},\ \bibinfo {pages} {195406} (\bibinfo {year} {2024}{\natexlab{a}})}\BibitemShut {NoStop}%
\bibitem [{\citenamefont {Yang}\ \emph {et~al.}(2024{\natexlab{b}})\citenamefont {Yang}, \citenamefont {Xu}, \citenamefont {Feng}, \citenamefont {Schindler}, \citenamefont {Xu}, \citenamefont {Bi}, \citenamefont {Bernevig}, \citenamefont {Tang},\ and\ \citenamefont {Liu}}]{yang_nc_2024}%
  \BibitemOpen
  \bibfield  {author} {\bibinfo {author} {\bibfnamefont {K.}~\bibnamefont {Yang}}, \bibinfo {author} {\bibfnamefont {Z.}~\bibnamefont {Xu}}, \bibinfo {author} {\bibfnamefont {Y.}~\bibnamefont {Feng}}, \bibinfo {author} {\bibfnamefont {F.}~\bibnamefont {Schindler}}, \bibinfo {author} {\bibfnamefont {Y.}~\bibnamefont {Xu}}, \bibinfo {author} {\bibfnamefont {Z.}~\bibnamefont {Bi}}, \bibinfo {author} {\bibfnamefont {B.~A.}\ \bibnamefont {Bernevig}}, \bibinfo {author} {\bibfnamefont {P.}~\bibnamefont {Tang}},\ and\ \bibinfo {author} {\bibfnamefont {C.-X.}\ \bibnamefont {Liu}},\ }\bibfield  {title} {\bibinfo {title} {Topological minibands and interaction driven quantum anomalous hall state in topological insulator based moiré heterostructures},\ }\href@noop {} {\bibfield  {journal} {\bibinfo  {journal} {Nature Communications}\ }\textbf {\bibinfo {volume} {15}} (\bibinfo {year} {2024}{\natexlab{b}})}\BibitemShut {NoStop}%
\bibitem [{\citenamefont {Bernevig}\ and\ \citenamefont {Zhang}(2006)}]{bernevig_quantum_2006}%
  \BibitemOpen
  \bibfield  {author} {\bibinfo {author} {\bibfnamefont {B.~A.}\ \bibnamefont {Bernevig}}\ and\ \bibinfo {author} {\bibfnamefont {S.-C.}\ \bibnamefont {Zhang}},\ }\bibfield  {title} {\bibinfo {title} {Quantum spin hall effect},\ }\href@noop {} {\bibfield  {journal} {\bibinfo  {journal} {Phys. Rev. Lett.}\ }\textbf {\bibinfo {volume} {96}},\ \bibinfo {pages} {106802} (\bibinfo {year} {2006})}\BibitemShut {NoStop}%
\bibitem [{\citenamefont {Wang}\ \emph {et~al.}(2013)\citenamefont {Wang}, \citenamefont {Weng}, \citenamefont {Wu}, \citenamefont {Dai},\ and\ \citenamefont {Fang}}]{wang_cd3as2_prb}%
  \BibitemOpen
  \bibfield  {author} {\bibinfo {author} {\bibfnamefont {Z.}~\bibnamefont {Wang}}, \bibinfo {author} {\bibfnamefont {H.}~\bibnamefont {Weng}}, \bibinfo {author} {\bibfnamefont {Q.}~\bibnamefont {Wu}}, \bibinfo {author} {\bibfnamefont {X.}~\bibnamefont {Dai}},\ and\ \bibinfo {author} {\bibfnamefont {Z.}~\bibnamefont {Fang}},\ }\bibfield  {title} {\bibinfo {title} {Three-dimensional dirac semimetal and quantum transport in cd${}_{3}$as${}_{2}$},\ }\href@noop {} {\bibfield  {journal} {\bibinfo  {journal} {Phys. Rev. B}\ }\textbf {\bibinfo {volume} {88}},\ \bibinfo {pages} {125427} (\bibinfo {year} {2013})}\BibitemShut {NoStop}%
\bibitem [{\citenamefont {Lygo}\ \emph {et~al.}(2023)\citenamefont {Lygo}, \citenamefont {Guo}, \citenamefont {Rashidi}, \citenamefont {Huang}, \citenamefont {Cuadros-Romero},\ and\ \citenamefont {Stemmer}}]{lygo_two-dimensional_2023}%
  \BibitemOpen
  \bibfield  {author} {\bibinfo {author} {\bibfnamefont {A.~C.}\ \bibnamefont {Lygo}}, \bibinfo {author} {\bibfnamefont {B.}~\bibnamefont {Guo}}, \bibinfo {author} {\bibfnamefont {A.}~\bibnamefont {Rashidi}}, \bibinfo {author} {\bibfnamefont {V.}~\bibnamefont {Huang}}, \bibinfo {author} {\bibfnamefont {P.}~\bibnamefont {Cuadros-Romero}},\ and\ \bibinfo {author} {\bibfnamefont {S.}~\bibnamefont {Stemmer}},\ }\bibfield  {title} {\bibinfo {title} {Two-{Dimensional} {Topological} {Insulator} {State} in {Cadmium} {Arsenide} {Thin} {Films}},\ }\href@noop {} {\bibfield  {journal} {\bibinfo  {journal} {Phys. Rev. Lett.}\ }\textbf {\bibinfo {volume} {130}},\ \bibinfo {pages} {046201} (\bibinfo {year} {2023})}\BibitemShut {NoStop}%
\bibitem [{\citenamefont {Guo}\ \emph {et~al.}(2023)\citenamefont {Guo}, \citenamefont {Miao}, \citenamefont {Huang}, \citenamefont {Lygo}, \citenamefont {Dai},\ and\ \citenamefont {Stemmer}}]{guo_zeeman_2023}%
  \BibitemOpen
  \bibfield  {author} {\bibinfo {author} {\bibfnamefont {B.}~\bibnamefont {Guo}}, \bibinfo {author} {\bibfnamefont {W.}~\bibnamefont {Miao}}, \bibinfo {author} {\bibfnamefont {V.}~\bibnamefont {Huang}}, \bibinfo {author} {\bibfnamefont {A.~C.}\ \bibnamefont {Lygo}}, \bibinfo {author} {\bibfnamefont {X.}~\bibnamefont {Dai}},\ and\ \bibinfo {author} {\bibfnamefont {S.}~\bibnamefont {Stemmer}},\ }\bibfield  {title} {\bibinfo {title} {Zeeman {Field}-{Induced} {Two}-{Dimensional} {Weyl} {Semimetal} {Phase} in {Cadmium} {Arsenide}},\ }\href@noop {} {\bibfield  {journal} {\bibinfo  {journal} {Phys. Rev. Lett.}\ }\textbf {\bibinfo {volume} {131}},\ \bibinfo {pages} {046601} (\bibinfo {year} {2023})}\BibitemShut {NoStop}%
\bibitem [{\citenamefont {Miao}\ \emph {et~al.}(2024)\citenamefont {Miao}, \citenamefont {Guo}, \citenamefont {Stemmer},\ and\ \citenamefont {Dai}}]{miao_engineering_2024}%
  \BibitemOpen
  \bibfield  {author} {\bibinfo {author} {\bibfnamefont {W.}~\bibnamefont {Miao}}, \bibinfo {author} {\bibfnamefont {B.}~\bibnamefont {Guo}}, \bibinfo {author} {\bibfnamefont {S.}~\bibnamefont {Stemmer}},\ and\ \bibinfo {author} {\bibfnamefont {X.}~\bibnamefont {Dai}},\ }\bibfield  {title} {\bibinfo {title} {Engineering the in-plane anomalous {Hall} effect in \ce{Cd3As2} thin films},\ }\href@noop {} {\bibfield  {journal} {\bibinfo  {journal} {Phys. Rev. B}\ }\textbf {\bibinfo {volume} {109}},\ \bibinfo {pages} {155408} (\bibinfo {year} {2024})}\BibitemShut {NoStop}%
\bibitem [{\citenamefont {König}\ \emph {et~al.}(2007)\citenamefont {König}, \citenamefont {Wiedmann}, \citenamefont {Brüne}, \citenamefont {Roth}, \citenamefont {Buhmann}, \citenamefont {Molenkamp}, \citenamefont {Qi},\ and\ \citenamefont {Zhang}}]{konig_qsh_2007}%
  \BibitemOpen
  \bibfield  {author} {\bibinfo {author} {\bibfnamefont {M.}~\bibnamefont {König}}, \bibinfo {author} {\bibfnamefont {S.}~\bibnamefont {Wiedmann}}, \bibinfo {author} {\bibfnamefont {C.}~\bibnamefont {Brüne}}, \bibinfo {author} {\bibfnamefont {A.}~\bibnamefont {Roth}}, \bibinfo {author} {\bibfnamefont {H.}~\bibnamefont {Buhmann}}, \bibinfo {author} {\bibfnamefont {L.~W.}\ \bibnamefont {Molenkamp}}, \bibinfo {author} {\bibfnamefont {X.-L.}\ \bibnamefont {Qi}},\ and\ \bibinfo {author} {\bibfnamefont {S.-C.}\ \bibnamefont {Zhang}},\ }\bibfield  {title} {\bibinfo {title} {Quantum spin hall insulator state in hgte quantum wells},\ }\href@noop {} {\bibfield  {journal} {\bibinfo  {journal} {Science}\ }\textbf {\bibinfo {volume} {318}},\ \bibinfo {pages} {766} (\bibinfo {year} {2007})}\BibitemShut {NoStop}%
\bibitem [{\citenamefont {Du}\ \emph {et~al.}(2015)\citenamefont {Du}, \citenamefont {Knez}, \citenamefont {Sullivan},\ and\ \citenamefont {Du}}]{du_qsh_2015}%
  \BibitemOpen
  \bibfield  {author} {\bibinfo {author} {\bibfnamefont {L.}~\bibnamefont {Du}}, \bibinfo {author} {\bibfnamefont {I.}~\bibnamefont {Knez}}, \bibinfo {author} {\bibfnamefont {G.}~\bibnamefont {Sullivan}},\ and\ \bibinfo {author} {\bibfnamefont {R.-R.}\ \bibnamefont {Du}},\ }\bibfield  {title} {\bibinfo {title} {Robust helical edge transport in gated $\mathrm{InAs}/\mathrm{GaSb}$ bilayers},\ }\href@noop {} {\bibfield  {journal} {\bibinfo  {journal} {Phys. Rev. Lett.}\ }\textbf {\bibinfo {volume} {114}},\ \bibinfo {pages} {096802} (\bibinfo {year} {2015})}\BibitemShut {NoStop}%
\bibitem [{\citenamefont {Tang}\ \emph {et~al.}(2017)\citenamefont {Tang}, \citenamefont {Zhang}, \citenamefont {Wong}, \citenamefont {Pedramrazi}, \citenamefont {Tsai}, \citenamefont {Jia}, \citenamefont {Moritz}, \citenamefont {Claassen}, \citenamefont {Ryu}, \citenamefont {Kahn}, \citenamefont {Jiang}, \citenamefont {Yan}, \citenamefont {Hashimoto}, \citenamefont {Lu}, \citenamefont {Moore}, \citenamefont {Hwang}, \citenamefont {Hwang}, \citenamefont {Hussain}, \citenamefont {Chen}, \citenamefont {Ugeda}, \citenamefont {Liu}, \citenamefont {Xie}, \citenamefont {Devereaux}, \citenamefont {Crommie}, \citenamefont {Mo},\ and\ \citenamefont {Shen}}]{tang_qsh_wte2_2017}%
  \BibitemOpen
  \bibfield  {author} {\bibinfo {author} {\bibfnamefont {S.}~\bibnamefont {Tang}}, \bibinfo {author} {\bibfnamefont {C.}~\bibnamefont {Zhang}}, \bibinfo {author} {\bibfnamefont {D.}~\bibnamefont {Wong}}, \bibinfo {author} {\bibfnamefont {Z.}~\bibnamefont {Pedramrazi}}, \bibinfo {author} {\bibfnamefont {H.-Z.}\ \bibnamefont {Tsai}}, \bibinfo {author} {\bibfnamefont {C.}~\bibnamefont {Jia}}, \bibinfo {author} {\bibfnamefont {B.}~\bibnamefont {Moritz}}, \bibinfo {author} {\bibfnamefont {M.}~\bibnamefont {Claassen}}, \bibinfo {author} {\bibfnamefont {H.}~\bibnamefont {Ryu}}, \bibinfo {author} {\bibfnamefont {S.}~\bibnamefont {Kahn}}, \bibinfo {author} {\bibfnamefont {J.}~\bibnamefont {Jiang}}, \bibinfo {author} {\bibfnamefont {H.}~\bibnamefont {Yan}}, \bibinfo {author} {\bibfnamefont {M.}~\bibnamefont {Hashimoto}}, \bibinfo {author} {\bibfnamefont {D.}~\bibnamefont {Lu}}, \bibinfo {author} {\bibfnamefont {R.~G.}\ \bibnamefont {Moore}}, \bibinfo {author} {\bibfnamefont {C.-C.}\ \bibnamefont {Hwang}}, \bibinfo
  {author} {\bibfnamefont {C.}~\bibnamefont {Hwang}}, \bibinfo {author} {\bibfnamefont {Z.}~\bibnamefont {Hussain}}, \bibinfo {author} {\bibfnamefont {Y.}~\bibnamefont {Chen}}, \bibinfo {author} {\bibfnamefont {M.~M.}\ \bibnamefont {Ugeda}}, \bibinfo {author} {\bibfnamefont {Z.}~\bibnamefont {Liu}}, \bibinfo {author} {\bibfnamefont {X.}~\bibnamefont {Xie}}, \bibinfo {author} {\bibfnamefont {T.~P.}\ \bibnamefont {Devereaux}}, \bibinfo {author} {\bibfnamefont {M.~F.}\ \bibnamefont {Crommie}}, \bibinfo {author} {\bibfnamefont {S.-K.}\ \bibnamefont {Mo}},\ and\ \bibinfo {author} {\bibfnamefont {Z.-X.}\ \bibnamefont {Shen}},\ }\bibfield  {title} {\bibinfo {title} {Quantum spin hall state in monolayer 1t’-wte2},\ }\href@noop {} {\bibfield  {journal} {\bibinfo  {journal} {Nature Physics}\ }\textbf {\bibinfo {volume} {13}},\ \bibinfo {pages} {683–687} (\bibinfo {year} {2017})}\BibitemShut {NoStop}%
\bibitem [{\citenamefont {Kane}\ and\ \citenamefont {Mele}(2005)}]{kane_mele_2005}%
  \BibitemOpen
  \bibfield  {author} {\bibinfo {author} {\bibfnamefont {C.~L.}\ \bibnamefont {Kane}}\ and\ \bibinfo {author} {\bibfnamefont {E.~J.}\ \bibnamefont {Mele}},\ }\bibfield  {title} {\bibinfo {title} {Quantum spin hall effect in graphene},\ }\href@noop {} {\bibfield  {journal} {\bibinfo  {journal} {Phys. Rev. Lett.}\ }\textbf {\bibinfo {volume} {95}},\ \bibinfo {pages} {226801} (\bibinfo {year} {2005})}\BibitemShut {NoStop}%
\bibitem [{\citenamefont {Bultinck}\ \emph {et~al.}(2020)\citenamefont {Bultinck}, \citenamefont {Khalaf}, \citenamefont {Liu}, \citenamefont {Chatterjee}, \citenamefont {Vishwanath},\ and\ \citenamefont {Zaletel}}]{nick_prx_2020}%
  \BibitemOpen
  \bibfield  {author} {\bibinfo {author} {\bibfnamefont {N.}~\bibnamefont {Bultinck}}, \bibinfo {author} {\bibfnamefont {E.}~\bibnamefont {Khalaf}}, \bibinfo {author} {\bibfnamefont {S.}~\bibnamefont {Liu}}, \bibinfo {author} {\bibfnamefont {S.}~\bibnamefont {Chatterjee}}, \bibinfo {author} {\bibfnamefont {A.}~\bibnamefont {Vishwanath}},\ and\ \bibinfo {author} {\bibfnamefont {M.~P.}\ \bibnamefont {Zaletel}},\ }\bibfield  {title} {\bibinfo {title} {Ground state and hidden symmetry of magic-angle graphene at even integer filling},\ }\href@noop {} {\bibfield  {journal} {\bibinfo  {journal} {Phys. Rev. X}\ }\textbf {\bibinfo {volume} {10}},\ \bibinfo {pages} {031034} (\bibinfo {year} {2020})}\BibitemShut {NoStop}%
\bibitem [{\citenamefont {Zhang}\ \emph {et~al.}(2020)\citenamefont {Zhang}, \citenamefont {Jiang}, \citenamefont {Wang},\ and\ \citenamefont {Zhang}}]{zhang_hf_2020}%
  \BibitemOpen
  \bibfield  {author} {\bibinfo {author} {\bibfnamefont {Y.}~\bibnamefont {Zhang}}, \bibinfo {author} {\bibfnamefont {K.}~\bibnamefont {Jiang}}, \bibinfo {author} {\bibfnamefont {Z.}~\bibnamefont {Wang}},\ and\ \bibinfo {author} {\bibfnamefont {F.}~\bibnamefont {Zhang}},\ }\bibfield  {title} {\bibinfo {title} {Correlated insulating phases of twisted bilayer graphene at commensurate filling fractions: A hartree-fock study},\ }\href@noop {} {\bibfield  {journal} {\bibinfo  {journal} {Phys. Rev. B}\ }\textbf {\bibinfo {volume} {102}},\ \bibinfo {pages} {035136} (\bibinfo {year} {2020})}\BibitemShut {NoStop}%
\bibitem [{\citenamefont {Zhang}\ \emph {et~al.}(2022)\citenamefont {Zhang}, \citenamefont {Lu},\ and\ \citenamefont {Liu}}]{zhang_citbg_prl_2022}%
  \BibitemOpen
  \bibfield  {author} {\bibinfo {author} {\bibfnamefont {S.}~\bibnamefont {Zhang}}, \bibinfo {author} {\bibfnamefont {X.}~\bibnamefont {Lu}},\ and\ \bibinfo {author} {\bibfnamefont {J.}~\bibnamefont {Liu}},\ }\bibfield  {title} {\bibinfo {title} {Correlated insulators, density wave states, and their nonlinear optical response in magic-angle twisted bilayer graphene},\ }\href {https://doi.org/10.1103/PhysRevLett.128.247402} {\bibfield  {journal} {\bibinfo  {journal} {Phys. Rev. Lett.}\ }\textbf {\bibinfo {volume} {128}},\ \bibinfo {pages} {247402} (\bibinfo {year} {2022})}\BibitemShut {NoStop}%
\bibitem [{\citenamefont {Tan}\ \emph {et~al.}(2024)\citenamefont {Tan}, \citenamefont {Reddy}, \citenamefont {Fu},\ and\ \citenamefont {Devakul}}]{trithep_2024}%
  \BibitemOpen
  \bibfield  {author} {\bibinfo {author} {\bibfnamefont {T.}~\bibnamefont {Tan}}, \bibinfo {author} {\bibfnamefont {A.~P.}\ \bibnamefont {Reddy}}, \bibinfo {author} {\bibfnamefont {L.}~\bibnamefont {Fu}},\ and\ \bibinfo {author} {\bibfnamefont {T.}~\bibnamefont {Devakul}},\ }\href@noop {} {\bibinfo {title} {Designing topology and fractionalization in narrow gap semiconductor films via electrostatic engineering}} (\bibinfo {year} {2024})\BibitemShut {NoStop}%
\bibitem [{\citenamefont {Tarnopolsky}\ \emph {et~al.}(2019)\citenamefont {Tarnopolsky}, \citenamefont {Kruchkov},\ and\ \citenamefont {Vishwanath}}]{ashvin_matbg_prl_2019}%
  \BibitemOpen
  \bibfield  {author} {\bibinfo {author} {\bibfnamefont {G.}~\bibnamefont {Tarnopolsky}}, \bibinfo {author} {\bibfnamefont {A.~J.}\ \bibnamefont {Kruchkov}},\ and\ \bibinfo {author} {\bibfnamefont {A.}~\bibnamefont {Vishwanath}},\ }\bibfield  {title} {\bibinfo {title} {Origin of magic angles in twisted bilayer graphene},\ }\href@noop {} {\bibfield  {journal} {\bibinfo  {journal} {Phys. Rev. Lett.}\ }\textbf {\bibinfo {volume} {122}},\ \bibinfo {pages} {106405} (\bibinfo {year} {2019})}\BibitemShut {NoStop}%
\bibitem [{\citenamefont {Liu}\ \emph {et~al.}(2021)\citenamefont {Liu}, \citenamefont {Khalaf}, \citenamefont {Lee},\ and\ \citenamefont {Vishwanath}}]{shang_nematic_tsm_2021}%
  \BibitemOpen
  \bibfield  {author} {\bibinfo {author} {\bibfnamefont {S.}~\bibnamefont {Liu}}, \bibinfo {author} {\bibfnamefont {E.}~\bibnamefont {Khalaf}}, \bibinfo {author} {\bibfnamefont {J.~Y.}\ \bibnamefont {Lee}},\ and\ \bibinfo {author} {\bibfnamefont {A.}~\bibnamefont {Vishwanath}},\ }\bibfield  {title} {\bibinfo {title} {Nematic topological semimetal and insulator in magic-angle bilayer graphene at charge neutrality},\ }\href@noop {} {\bibfield  {journal} {\bibinfo  {journal} {Phys. Rev. Res.}\ }\textbf {\bibinfo {volume} {3}},\ \bibinfo {pages} {013033} (\bibinfo {year} {2021})}\BibitemShut {NoStop}%
\bibitem [{\citenamefont {Shi}\ \emph {et~al.}(2024)\citenamefont {Shi}, \citenamefont {Miao},\ and\ \citenamefont {Dai}}]{shi_epc_tbg_2024}%
  \BibitemOpen
  \bibfield  {author} {\bibinfo {author} {\bibfnamefont {H.}~\bibnamefont {Shi}}, \bibinfo {author} {\bibfnamefont {W.}~\bibnamefont {Miao}},\ and\ \bibinfo {author} {\bibfnamefont {X.}~\bibnamefont {Dai}},\ }\href@noop {} {\bibinfo {title} {Moiré optical phonons dancing with heavy electrons in magic-angle twisted bilayer graphene}} (\bibinfo {year} {2024})\BibitemShut {NoStop}%
\bibitem [{\citenamefont {Wang}\ \emph {et~al.}(2024{\natexlab{c}})\citenamefont {Wang}, \citenamefont {Zhou}, \citenamefont {Lian},\ and\ \citenamefont {Song}}]{wang_epc_tbg_2024}%
  \BibitemOpen
  \bibfield  {author} {\bibinfo {author} {\bibfnamefont {Y.-J.}\ \bibnamefont {Wang}}, \bibinfo {author} {\bibfnamefont {G.-D.}\ \bibnamefont {Zhou}}, \bibinfo {author} {\bibfnamefont {B.}~\bibnamefont {Lian}},\ and\ \bibinfo {author} {\bibfnamefont {Z.-D.}\ \bibnamefont {Song}},\ }\href@noop {} {\bibinfo {title} {Electron phonon coupling in the topological heavy fermion model of twisted bilayer graphene}} (\bibinfo {year} {2024}{\natexlab{c}})\BibitemShut {NoStop}%
\bibitem [{\citenamefont {Sheng}\ \emph {et~al.}(2024)\citenamefont {Sheng}, \citenamefont {Reddy}, \citenamefont {Abouelkomsan}, \citenamefont {Bergholtz},\ and\ \citenamefont {Fu}}]{sheng_qah_prl_2024}%
  \BibitemOpen
  \bibfield  {author} {\bibinfo {author} {\bibfnamefont {D.~N.}\ \bibnamefont {Sheng}}, \bibinfo {author} {\bibfnamefont {A.~P.}\ \bibnamefont {Reddy}}, \bibinfo {author} {\bibfnamefont {A.}~\bibnamefont {Abouelkomsan}}, \bibinfo {author} {\bibfnamefont {E.~J.}\ \bibnamefont {Bergholtz}},\ and\ \bibinfo {author} {\bibfnamefont {L.}~\bibnamefont {Fu}},\ }\bibfield  {title} {\bibinfo {title} {Quantum anomalous hall crystal at fractional filling of moir\'e superlattices},\ }\href@noop {} {\bibfield  {journal} {\bibinfo  {journal} {Phys. Rev. Lett.}\ }\textbf {\bibinfo {volume} {133}},\ \bibinfo {pages} {066601} (\bibinfo {year} {2024})}\BibitemShut {NoStop}%
\bibitem [{\citenamefont {Zeng}\ \emph {et~al.}(2024{\natexlab{b}})\citenamefont {Zeng}, \citenamefont {Guerci}, \citenamefont {Cr\'epel}, \citenamefont {Millis},\ and\ \citenamefont {Cano}}]{zeng_topo_ahc_2024}%
  \BibitemOpen
  \bibfield  {author} {\bibinfo {author} {\bibfnamefont {Y.}~\bibnamefont {Zeng}}, \bibinfo {author} {\bibfnamefont {D.}~\bibnamefont {Guerci}}, \bibinfo {author} {\bibfnamefont {V.}~\bibnamefont {Cr\'epel}}, \bibinfo {author} {\bibfnamefont {A.~J.}\ \bibnamefont {Millis}},\ and\ \bibinfo {author} {\bibfnamefont {J.}~\bibnamefont {Cano}},\ }\bibfield  {title} {\bibinfo {title} {Sublattice structure and topology in spontaneously crystallized electronic states},\ }\href@noop {} {\bibfield  {journal} {\bibinfo  {journal} {Phys. Rev. Lett.}\ }\textbf {\bibinfo {volume} {132}},\ \bibinfo {pages} {236601} (\bibinfo {year} {2024}{\natexlab{b}})}\BibitemShut {NoStop}%
\bibitem [{\citenamefont {Dong}\ \emph {et~al.}(2023)\citenamefont {Dong}, \citenamefont {Wang}, \citenamefont {Wang}, \citenamefont {Soejima}, \citenamefont {Zaletel}, \citenamefont {Vishwanath},\ and\ \citenamefont {Parker}}]{dong_ahc_2023}%
  \BibitemOpen
  \bibfield  {author} {\bibinfo {author} {\bibfnamefont {J.}~\bibnamefont {Dong}}, \bibinfo {author} {\bibfnamefont {T.}~\bibnamefont {Wang}}, \bibinfo {author} {\bibfnamefont {T.}~\bibnamefont {Wang}}, \bibinfo {author} {\bibfnamefont {T.}~\bibnamefont {Soejima}}, \bibinfo {author} {\bibfnamefont {M.~P.}\ \bibnamefont {Zaletel}}, \bibinfo {author} {\bibfnamefont {A.}~\bibnamefont {Vishwanath}},\ and\ \bibinfo {author} {\bibfnamefont {D.~E.}\ \bibnamefont {Parker}},\ }\href@noop {} {\bibinfo {title} {Anomalous hall crystals in rhombohedral multilayer graphene i: Interaction-driven chern bands and fractional quantum hall states at zero magnetic field}} (\bibinfo {year} {2023})\BibitemShut {NoStop}%
\bibitem [{\citenamefont {Dong}\ \emph {et~al.}(2024)\citenamefont {Dong}, \citenamefont {Patri},\ and\ \citenamefont {Senthil}}]{dong_ahc_senthil_2024}%
  \BibitemOpen
  \bibfield  {author} {\bibinfo {author} {\bibfnamefont {Z.}~\bibnamefont {Dong}}, \bibinfo {author} {\bibfnamefont {A.~S.}\ \bibnamefont {Patri}},\ and\ \bibinfo {author} {\bibfnamefont {T.}~\bibnamefont {Senthil}},\ }\href@noop {} {\bibinfo {title} {Stability of anomalous hall crystals in multilayer rhombohedral graphene}} (\bibinfo {year} {2024})\BibitemShut {NoStop}%
\bibitem [{\citenamefont {Tan}\ and\ \citenamefont {Devakul}(2024)}]{tan_parent_bc_2024}%
  \BibitemOpen
  \bibfield  {author} {\bibinfo {author} {\bibfnamefont {T.}~\bibnamefont {Tan}}\ and\ \bibinfo {author} {\bibfnamefont {T.}~\bibnamefont {Devakul}},\ }\href@noop {} {\bibinfo {title} {Parent berry curvature and the ideal anomalous hall crystal}} (\bibinfo {year} {2024})\BibitemShut {NoStop}%
\bibitem [{\citenamefont {Kwan}\ \emph {et~al.}(2021)\citenamefont {Kwan}, \citenamefont {Devakul}, \citenamefont {Sondhi},\ and\ \citenamefont {Parameswaran}}]{kwan_prb_2021}%
  \BibitemOpen
  \bibfield  {author} {\bibinfo {author} {\bibfnamefont {Y.~H.}\ \bibnamefont {Kwan}}, \bibinfo {author} {\bibfnamefont {T.}~\bibnamefont {Devakul}}, \bibinfo {author} {\bibfnamefont {S.~L.}\ \bibnamefont {Sondhi}},\ and\ \bibinfo {author} {\bibfnamefont {S.~A.}\ \bibnamefont {Parameswaran}},\ }\bibfield  {title} {\bibinfo {title} {Theory of competing excitonic orders in insulating ${\mathrm{wte}}_{2}$ monolayers},\ }\href@noop {} {\bibfield  {journal} {\bibinfo  {journal} {Phys. Rev. B}\ }\textbf {\bibinfo {volume} {104}},\ \bibinfo {pages} {125133} (\bibinfo {year} {2021})}\BibitemShut {NoStop}%
\bibitem [{\citenamefont {Zhu}\ \emph {et~al.}(2024)\citenamefont {Zhu}, \citenamefont {Cookmeyer},\ and\ \citenamefont {Das~Sarma}}]{jihang_dw_2024}%
  \BibitemOpen
  \bibfield  {author} {\bibinfo {author} {\bibfnamefont {J.}~\bibnamefont {Zhu}}, \bibinfo {author} {\bibfnamefont {T.}~\bibnamefont {Cookmeyer}},\ and\ \bibinfo {author} {\bibfnamefont {S.}~\bibnamefont {Das~Sarma}},\ }\bibfield  {title} {\bibinfo {title} {Pseudospin density wave instability in two-dimensional electron bilayers},\ }\href@noop {} {\bibfield  {journal} {\bibinfo  {journal} {Phys. Rev. B}\ }\textbf {\bibinfo {volume} {110}},\ \bibinfo {pages} {054405} (\bibinfo {year} {2024})}\BibitemShut {NoStop}%
\bibitem [{\citenamefont {Tange}(2024)}]{gnu_parallel}%
  \BibitemOpen
  \bibfield  {author} {\bibinfo {author} {\bibfnamefont {O.}~\bibnamefont {Tange}},\ }\href {https://doi.org/10.5281/ZENODO.11043435} {\bibinfo {title} {Gnu parallel 20240422 ('børsen')}} (\bibinfo {year} {2024})\BibitemShut {NoStop}%
\end{thebibliography}%
\end{document}